\documentclass[11pt,a4paper]{article}
\pdfoutput=1
\usepackage[T1]{fontenc}
\usepackage{lmodern,microtype,booktabs,tikz}
\usepackage{jheppub} 
\hypersetup{colorlinks=true,linkcolor=blue,citecolor=blue,urlcolor=blue,pdftitle={Actionless Branes on Axion Wormholes},pdfauthor={A. Ramesh Chandra}}
\newcommand{\arx}[1]{\href{https://arxiv.org/abs/#1}{arXiv:#1}}
\newcommand{\dd}{\mathrm{d}}
\newcommand{\bd}{\mathbf d}   
\newcommand{\bz}{\mathbf z}   
\newcommand{\Imm}{\operatorname{Im}}
\newcommand{\Ree}{\operatorname{Re}}
\usepackage{letltxmacro}
\AtBeginDocument{\LetLtxMacro\NATcite\cite
 \renewcommand{\cite}[2][]{\NATcite{#2}}}
\begin{document}
\title{Actionless Branes on Axion Wormholes}
\author{A. Ramesh Chandra}
\emailAdd{ramesh.chandra@icts.res.in}
\affiliation{International Centre for Theoretical Sciences,\\
Tata Institute of Fundamental Research,\\
Shivakote, Bengaluru 560089, India}

\abstract{
Can a Euclidean axion wormhole extend to infinity with every instanton still
suppressed? The imaginary distance bound says it cannot: some instanton
must become actionless first. Calabi--Yau compactifications appear to
violate the bound: several axions share the field excursion, and the
instantons charged under any one of them stay suppressed all the way. We
show that the wormhole can hide from these instantons but not from their
bound states: misaligned central charges make a bound state cheaper than
its constituents, while the axions act on every constituent. On the type
IIB wormhole supported by two axions of the Neveu--Schwarz two-form, a
fivebrane
carrying string charge becomes actionless at large volume within two
thirds of the way to infinity, for either relative sign of the axions
and however they share the excursion. In S-dual variables it is a
D5-brane with magnetic flux for one sign and a D5--NS5 bound state for the other. Adding more
axions does not help: the same branes become actionless on every Calabi--Yau
threefold, because what counts is the degree of homogeneity of the
saxion sector, not the number of axions. These bound states are what the
convex-hull axion weak gravity conjecture requires.
}
\maketitle

\section{Introduction}

Axions have played a central role in the study of Euclidean wormholes from the very beginnings of the subject~\cite{Hawking1988,Giddings1988,Coleman1988}. The seminal construction of Giddings and Strominger demonstrated that axion flux can support a smooth Euclidean wormhole connecting two asymptotically flat regions~\cite{Giddings1988}. If they contribute to the gravitational path integral, Euclidean wormholes could have far-reaching consequences: they can render low-energy couplings stochastic~\cite{Coleman1988}, induce violations of global symmetries, and, in anti-de Sitter space, pose a puzzle for the factorization of boundary partition functions~\cite{MaldacenaMaoz,ArkaniHamed2007,Held2026}. Axion-supported wormholes also arise in the low-energy effective theories of string compactifications~\cite{ArkaniHamed2007,Hebecker2018,Martucci2024}. Whether these solutions survive as legitimate saddles of the full theory is therefore a fundamental question.

One feature of axion wormholes makes the question sharp. The throat
can be made arbitrarily large by increasing the flux, but the distance
traveled by the scalar fields cannot. From a reflection-symmetric
throat to either asymptotically flat end in $D$ spacetime dimensions,
the scalars travel a fixed target-space proper distance, the
excursion~\cite{Giddings1988,ArkaniHamed2007,Maldacena2026}
\begin{equation}
 \tau_{\rm end}=\frac{\pi}{2}\sqrt{\frac{2(D-1)}{D-2}} .
 \label{eq:idb}
\end{equation}
Here and throughout, scalar field-space distances are measured in
reduced Planck units. In the scalar description the axions are
imaginary on the solution, so the wormhole is a path of fixed length
into complexified field space. Maldacena, Maloney and McPeak proposed
that this length is a bound. When the massless couplings of a
consistent theory are continued away from the real slice, some effect
must invalidate the low-energy approximation at or before imaginary
distance $\tau_{\rm end}$. This is the imaginary distance bound
(IDB)~\cite{Maldacena2026}.

In the known examples the effect is an instanton that loses its
exponential suppression. An instanton of charge $q$, whose action 
is $U$ when the axion vanishes, has weight $\mathcal P\,e^{-U-2\pi iqa}$ 
at real axion $a$, where the prefactor $\mathcal P$ comes from its fluctuations and zero modes.
At $a=iT$ the exponential becomes $e^{-U+2\pi qT}$, and for one
orientation the axion offsets part of the cost $U$. In ten-dimensional type
IIB with Ramond--Ramond (RR) axion $C_0=iT$, a D-instanton has action
$2\pi(g_s^{-1}-|T|)$. It vanishes at $|T|=g_s^{-1}$, where the
coefficient of the $R^4$ interaction
diverges~\cite{Maldacena2026,Green1997}. Requiring an actionless
instanton within a fixed distance bounds the action per unit charge,
which makes the IDB a close relative of the axion weak gravity
conjecture~\cite[App.~A]{AMNV2007,Montero2015,Maldacena2026}. The same ratio enters the
positivity of the gravitational path integral~\cite{DiUbaldo2026} and decides whether inserting a charged defect lowers the action of a wormhole~\cite{ArkaniHamed2007,Lin2026}.

Several axions pose a sharper test~\cite[Sec.~5.2.1]{Maldacena2026}.
They can share the excursion, and an instanton charged under one of
them sees only part of it. The four-dimensional wormhole of a
Calabi--Yau compactification is the basic
example~\cite{Tamvakis,Bergshoeff2004,Bergshoeff2005,ArkaniHamed2007}. A fundamental string
wrapping a curve feels the internal Neveu--Schwarz (NS) two-form. An NS
fivebrane (NS5) wrapping the threefold feels the axion dual to the
spacetime two-form. These are the elementary instantons, each charged
under one axion. If the two axions share the excursion, both stay
suppressed all the way to the asymptotic region, and any collection of
instantons whose actions simply add fares no better. With hundreds of
axions, each might move only a short way, and the low-energy theory
would show no sign of trouble. The IDB then demands an instanton that is not elementary. A simpler
setting shows what one looks like: on a product of two two-tori, a
single string winding diagonally across both loses its action while
strings wrapped on either torus stay suppressed. A similar breakdown is
expected for the Calabi--Yau wormhole, but none has been
found~\cite{Maldacena2026}. Which object plays this role on a curved
threefold, and does one exist however the excursion is shared?

Two recent works show that the difficulty is structural. Licciardello
and Martucci study supersymmetric axion sectors whose K\"ahler
potential is minus the logarithm of a homogeneous function of the
saxions. On the wormhole along which all saxions grow in proportion,
the mutually supersymmetric instantons charged under these axions
become actionless only if the degree of homogeneity is at most
three~\cite{Licciardello2026}. Etheredge, Reece, Rudelius and Tudball
show that the same degree, rather than the number of axions, sets the
smallest charge per unit action of such instantons, and that degrees
add when sectors are combined~\cite{Etheredge2026}. The dilaton and the
K\"ahler moduli of a Calabi--Yau threefold, on the slice where the RR
axions vanish, form such a sector of degree four: one from the dilaton
and three from the cubic volume, irrespective of the number of
K\"ahler moduli. The failure of the string and the NS5 is
therefore not cured by adding more axions. For degree above three,
Ref.~\cite{Licciardello2026} argues that instantons outside the
mutually supersymmetric class must enforce the imaginary distance
bound. It predicts that these instantons are lighter than their BPS
counterparts, derives a bound on their action relative to the
wormhole, and calls for a test in string theory. Its degree-four
example has precisely the one-plus-three homogeneity structure of
the Calabi--Yau sector considered here. Reference.~\cite{Etheredge2026}
also asks whether its bounds on supersymmetric instantons extend to
convex-hull conditions over the entire charge lattice. The same difficulty has recently been noted for the D-instantons of a
toroidal orbifold that are T-dual to BPS black holes~\cite{Delgado:2026bhp}.
Those charged under all four RR axions lie in a sector of degree four and
their actions simply add, so by the argument of Ref.~\cite{Licciardello2026} only
instantons outside the mutually supersymmetric class could enforce the
bound there.

We give a concrete string-theoretic realization of the additional
instantons required by Ref.~\cite{Licciardello2026}. On the wormhole studied there, along which the two imaginary axions have
the same sign, a D5--NS5 bound state becomes actionless before the
asymptotic end, although the mutually supersymmetric elementary
instantons remain suppressed. On its reflection, the opposite-sign
wormhole, a fivebrane carrying magnetic flux also becomes actionless. We further establish the
convex-hull bound for the K\"ahler axions of any Calabi--Yau
threefold directly from the cubic volume, without invoking the
higher-dimensional weak gravity conjecture used in
Ref.~\cite{Etheredge2026}.

The basic mechanism is simple. Write the real part of the action
of a Euclidean brane on the wormhole as a wrapping cost minus the
work done on it by the imaginary axions, $\mathrm{Re}\,I=U-W$, where $U$ is
its action with the axions set to zero. The brane loses its
exponential suppression when $W/U$ reaches one. When the axions
couple linearly to the charges, this ratio is the charge per unit
cost contracted with the axion displacement, so the work done on
a bound state is the sum of the work on its constituents. The
wrapping cost, however, is not additive. At large volume it is
the modulus of a central charge, and a bound state can cost less than its separated constituents because their central charges have different phases. It can therefore lose its entire action while each
constituent remains suppressed. These bound states are BPS
instantons of the full theory, but their constituents are not
mutually supersymmetric, placing them outside the class considered
above.\footnote{References.~\cite{Licciardello2026,Etheredge2026} reserve
``BPS'' for instantons that are mutually supersymmetric with the
elementary ones; in that usage our bound states are non-BPS. We call
a brane BPS when its cost saturates the central-charge bound
(Sec.~\ref{sec:criterion}).}

\emph{The two-axion wormhole.} The instanton that enforces the bound is a
fivebrane carrying string charge. It reaches its suppression wall, the point where
its action vanishes, at
\begin{equation}
 \tau_*=\frac{\pi}{p_s+\sqrt3\,p_v},\qquad
 \frac{1}{\sqrt3}\le\frac{\tau_*}{\tau_{\rm end}}<\frac23 ,
 \label{eq:headline}
\end{equation}
where $p_s$ and $p_v$, with $p_s^2+p_v^2=1$, are the speeds in the
external and internal axion--saxion planes, so that their ratio says how
the excursion is shared, and $\tau_{\rm end}=\pi\sqrt3/2$. This holds for
every regular division of the excursion, one that keeps the saxions
finite to the end, and for either relative sign of the two axions. The
sign decides the brane. For one sign it is, in the S-dual description,
a D5 that carries string
charge through a real magnetic flux. For the other, the charges on which the axions do positive work are mutually supersymmetric and binding them gains nothing; instead, the
continued two-form itself supplies an imaginary worldvolume field to a
bound state of a D5 and an NS5, and the field does additional work on
it. The two branes have different actions but the same wall. 

We observe
that this wall lies where the wormhole has made a quarter turn along a root of
the full scalar manifold, a root whose axion the wormhole leaves at zero
(Sec.~\ref{sec:geometry}). A third
brane, a D5 of the original variables, on which the axion makes the worldvolume field complex,
loses its suppression later, without its action vanishing. The fluxed D5 can also give an order-one correction to an axion kinetic term before its wall is reached.

\emph{Many axions.} When all the curve areas grow in proportion, so that
the K\"ahler form moves along a ray in the K\"ahler cone, the same
branes have the same walls on every threefold. When the shape of the threefold changes, two statements
hold. The wrapped strings of both orientations satisfy a convex-hull
condition for the K\"ahler axions, with exactly the radius that degree
three requires.
With the dilaton the radius is that of degree four, below what a wall
before the end needs, so the elementary instantons cannot guarantee a
wall for any number of axions. And both fivebranes are BPS, so they
persist when the shape changes slightly: near the ray they still cut the
wormhole off, for both signs. Far from the ray the question remains open.

What does this mean for the wormhole? With the boundary conditions of Ref.~\cite{Maldacena2026}, where the scalars are fixed at the ends, the low-energy theory fails before the end is reached, so the wormhole of the two-derivative theory is not a trustworthy saddle with these boundary values. With the charges fixed instead, the same branes say something different. For one relative sign, on every regular division, inserting the D5 with flux lowers the action of the wormhole~\cite{Lin2026}. For both signs, these branes are precisely the kinds of objects required by the convex-hull form of the axion weak gravity conjecture. They carry charge under axions that the wormhole leaves at zero, so a theory truncated to the axions that move cannot see their full charge vectors.

Section~\ref{sec:trajectory} sets up the wormhole, the two boundary
conditions under which it is studied, specifies the regimes in which it can be trusted, and the test for
an unsuppressed instanton. Section~\ref{sec:fivebranes} takes in turn
the real, imaginary and complex worldvolume fields through which the
internal axion can act on a fivebrane, finds the walls for every division of the
excursion, reads them off the geometry of the scalar manifold, and draws
their consequences for both boundary conditions. Section~\ref{sec:many} turns to
many axions, and Sec.~\ref{sec:discussion} discusses what the results
mean and what remains open. The appendices contain a bound on the work
of magnetic flux (\ref{app:flux-bounds}) and the proof of the hull
inequality (\ref{app:hull}).

\section{The saddle and its instantons}\label{sec:trajectory}

\subsection{The two-axion wormhole}\label{sec:wormhole}

We begin with the two-axion wormhole in the tree-level,
two-derivative supergravity approximation. We compactify type IIB
supergravity on a smooth projective Calabi--Yau threefold $X$, with
$h^{1,0}=h^{2,0}=0$, no background fluxes or orientifold
projection. In this approximation, the RR potentials
can be set to zero. Both axions supporting the wormhole descend
from the NS two-form $B_2$.

The spacetime component $B_{\mu\nu}$ is dual in four dimensions to
an axion $a$. Its non-periodic scalar partner, called a saxion, is
the four-dimensional dilaton
$s=e^{-2\Phi_4}=V/g_s^2$, where $V$ is the
string-frame volume of $X$ and $g_s$ is the ten-dimensional string
coupling.\footnote{We use $\ell_s=2\pi\sqrt{\alpha'}$, measure the
K\"ahler form $J$ and $B_{\rm int}$ in units of $\ell_s^2$, and
define $V=\operatorname{Vol}_{\rm str}(X)/\ell_s^6$.}
The second axion, $b$, arises from the internal components of $B_2$.
For simplicity, we keep the relative sizes of the internal cycles
fixed, allowing only the overall size of $X$ to vary. We also
restrict the internal two-form to the same direction as the K\"ahler
form. Thus, we write
\begin{equation}
 J=vH,\qquad B_{\rm int}=bH,\qquad
 V=\frac{\kappa v^3}{6},\qquad
 \kappa=\int_X H^3 ,
 \label{eq:ray}
\end{equation}
where $H$ is a fixed K\"ahler class, chosen to be primitive and
integral so that $b$ has period one. The saxion of $b$ is the area
modulus $v$. This restriction allows the shape of $X$ to remain
fixed even when $h^{1,1}>1$, and it is a consistent truncation
(Sec.~\ref{sec:ray}). To fix the axion normalizations, an NS5-brane wrapping $X$ has
semiclassical weight $e^{-2\pi s-2\pi ia}$, while a fundamental
string wrapping a curve $\Sigma$ of degree $n=\int_\Sigma H$ has
weight $e^{-2\pi nv-2\pi inb}$, each for one orientation and up
to a prefactor. Both axions have period one. The Euclidean action for real $a,b$ is
\begin{equation}
I_{\rm bulk}=\frac{1}{16\pi G_4}\int\dd^4x\,\sqrt g\,
 \biggl[-R+\frac{(\nabla s)^2+(\nabla a)^2}{2s^2} +3\,\frac{(\nabla v)^2+(\nabla b)^2}{2v^2}\biggr],
 \label{eq:bulk}
\end{equation}
so the scalars move on two hyperbolic planes of radii $1$ and $\sqrt3$,
\begin{equation}
 \dd\ell^2=\frac{\dd s^2+\dd a^2}{s^2}
 +3\frac{\dd v^2+\dd b^2}{v^2}.
 \label{eq:metric}
\end{equation}

The Euclidean wormhole is supported by real three-form fluxes dual to
the axions $a$ and $b$. In Euclidean signature the duality introduces a
factor of $i$, so in the scalar variables the axions are imaginary on
the solution. The same local solution appears with two different
boundary conditions, which differ in what is held fixed at the ends.
With \emph{fixed charges} the conserved fluxes $n_a,n_b$ are given. As Witten
has emphasized~\cite{Witten}, the scalar profiles are then only a way of
solving the classical equations: the duality involves a Poisson
resummation over flux sectors, and a wormhole of fixed charge has no
semiclassical description in terms of the scalars. With \emph{fixed
boundary values} the scalars are given at the two ends, at
values continued away from the real slice. This is the setting of
Ref.~\cite{Maldacena2026}. The asymptotic values of the scalars are couplings, and the imaginary distance bound limits how far the low-energy
approximation remains valid under this continuation. The two
are related by a Legendre transform at the level of the saddle and by a
Fourier transform over flux sectors in the full path integral. We work
with fixed boundary values, and return to fixed charges only in
Sec.~\ref{sec:consequences}.

The wormhole follows a timelike geodesic of the complexified scalar
metric~\eqref{eq:metric}, with $\dd\ell^2=-\dd\tau^2$. Here $\tau$ is the
target-space proper distance from the throat, in the normalization of
Ref.~\cite{Maldacena2026}, measured along one mouth: the half of the
wormhole that runs from the throat to one asymptotic end.
The Einstein constraint ties this distance to the
size of the wormhole.
For the spacetime metric
$\dd w^2+r(w)^2\dd\Omega_3^2$ with throat radius $r_0$ it gives
$(\dd r/\dd w)^2=1-r_0^4/r^4$ and
$r^3\,\dd\tau/\dd w=2\sqrt3\,r_0^2$~\cite{Tamvakis,ArkaniHamed2007},
and eliminating $w$,
\begin{equation}
 r^2=r_0^2\,\sec\frac{\tau}{\sqrt3}.
 \label{eq:radius}
\end{equation}
Spatial infinity is therefore reached at
\begin{equation}
 \tau_{\rm end}=\frac{\pi\sqrt3}{2}
 \label{eq:end}
\end{equation}
independent of $r_0$. This fixed excursion can be shared between the two axion--saxion pairs. We set the real parts of the axions to zero; at leading order they
change only the phase of the weights. On the mouth with $\Imm a>0$
the geodesic through a throat with saxions $s_0,v_0$ is
\begin{equation}
 \begin{aligned}
 s &= s_0\sec u_s, &\quad a &= i s_0\tan u_s, &\quad u_s&=p_s\tau,\\
 v &= v_0\sec u_v, &\quad b &= i\epsilon v_0\tan u_v, &\quad
 u_v&=\frac{p_v\tau}{\sqrt3}.
 \end{aligned}
 \label{eq:trajectory}
\end{equation}
Here $p_s$ and $p_v$ are the speeds in the two hyperbolic planes, with
$p_s^2+p_v^2=1$, and each angle $u_i$ is the distance travelled in its
hyperbolic plane divided by its radius. The sign $\epsilon=\pm1$ is
the \emph{relative sign} of the two axions. On this mouth the
\emph{same-sign} wormhole has $(\Imm a,\Imm b)$ of signs $(+,+)$ and
$\epsilon=+1$, and the \emph{opposite-sign} wormhole has $(+,-)$ and
$\epsilon=-1$; the two are distinguished by the sign of $n_an_b$, up to
orientation conventions. Each saxion diverges when its angle reaches
$\pi/2$. Both stay finite through $\tau_{\rm end}$ precisely when
$0<p_s<1/\sqrt3$, the \emph{regular} divisions of the
excursion~\cite{ArkaniHamed2007,Chiodaroli2009}. The equal-angle
choice $p_s=1/2$ gives $u_s=u_v=\tau/2$, which end at
$\pi\sqrt3/4<\pi/2$. The conserved fluxes are in the ratio
\begin{equation}
\left|\frac{n_b}{n_a}\right| = \sqrt3\,\frac{s_0}{v_0}\,\frac{p_v}{p_s},
\label{eq:fluxratio}
\end{equation}
set by the throat moduli and by how the excursion is divided.

With a single axion the wormhole is essentially unique, and the bound
is a statement about one instanton. With two it is a family, labelled
by the division $p_s$ and the relative sign $\epsilon$. Every member
reaches its end at the same $\tau_{\rm end}$, but along a different
direction in field space, and how fast an instanton loses its action
depends on that direction. The bound must therefore hold member by
member, and which brane enforces it, and where, can change from one
member to the next.

\subsection{Where the saddle can be trusted}\label{sec:frames}

The wormhole solves the two-derivative theory~\eqref{eq:bulk}. It is a saddle of string theory when this effective theory is accurate along the whole solution, which requires two things.
First, curvatures must be small. Multiplying both fluxes by a common factor
$\lambda$ enlarges the throat by $\sqrt\lambda$ in Planck units at
fixed moduli and fixed excursion, so the throat can be made large
compared with the Planck and compactification scales. Second,
Eq.~\eqref{eq:metric} must be the scalar metric to good accuracy all
along the path. Its corrections are either perturbative, suppressed by
inverse powers of the saxions, or instantons, suppressed by their
actions. They are controlled when both kinds of string wrapped on a
curve of unit degree are heavy,
\begin{equation}
 v\gg1,\qquad t\equiv\frac{v}{g_s}=\sqrt{\frac{6s}{\kappa v}}\gg1 ,
 \label{eq:t}
\end{equation}
the fundamental string costing $2\pi v$ and the D-string $2\pi t$. We also need $g_s^2=\kappa v^3/(6s)$ far from one. For $g_s\ll1$, that is
$s\gg v^3$, the corrections are those of weakly coupled type IIB at
large volume. For $g_s\gg1$, that is $v\ll s\ll v^3$, they are those
of the S-dual string, whose coupling is $1/g_s$ and whose K\"ahler
modulus is $t$. The four-dimensional action~\eqref{eq:bulk} is the same
in both cases, since it does not refer to a ten-dimensional
description. Only the expansion that controls its corrections changes.
For $g_s\sim1$ no duality frame is weakly coupled and we make no claim.
These conditions only need to be checked at the throat. Along a regular
wormhole the saxions grow, while $t$ and $g_s$ change by factors of
order one unless the division is close to an edge of the regular
range, so a throat well inside either regime gives a path inside
it.\footnote{The geodesic gives $t^2/t_0^2=\cos u_v/\cos u_s$ and
$g_s^2/g_{s,0}^2=\cos u_s/\cos^3u_v$. At equal angles $t=t_0$ is
constant and $g_s=g_{s,0}\sec(\tau/2)$.}

We name the two regimes after the potentials that support the wormhole
in the weakly coupled description. In the \emph{NS--NS regime},
$s_0\gg v_0^3$, these are $B_2$ and its dual. In the \emph{R--R
regime}, $v_0\ll s_0\ll v_0^3$, we use the S-dual variables, which we
mark with a tilde,
\begin{equation}
 \tilde g_s=\frac1{g_s},\quad \tilde J=tH,\quad \tilde B_2=0,\quad \tilde C_2=B_2,\quad
 \int_X\tilde C_6=a ,
 \label{eq:tilde}
\end{equation}
with $C_0=0$~\cite{Alexandrov2010,Alexandrov2021}. The wormhole is then
an RR-axion wormhole of weakly coupled type IIB, and D-branes feel it
through their Wess--Zumino couplings. 

On a wormhole in either regime, every correction we have listed is small at the throat, and the elementary instantons considered next
remain suppressed to the end. Throughout, the complex structure and the
RR axions are held fixed, the throat is reflection symmetric, and brane
actions are evaluated at leading order at large volume in one of the
two regimes. Axion wormholes are perturbatively stable at fixed
charge~\cite{Loges2022,Jonas2024,Hertog2024,Marolf2025}; with fixed
boundary values the throat size is instead a flat
direction~\cite[Sec.~2.1]{Maldacena2026}. Our question about this
saddle is a different one: whether every instanton of the theory, not
only the elementary ones, remains suppressed along it.

\subsection{Cost, work and the wall}\label{sec:criterion}

We need a test for when an instanton stops being suppressed on the
wormhole. Consider a Euclidean brane of charge $\gamma$ inserted at
distance $\tau$ from the throat. Its \emph{cost} $U_\gamma$ is its
action at the local saxions with the axions set to zero. The axions
lower the real part of the action, and we call the amount they remove
the \emph{work}, $W_\gamma\equiv U_\gamma-\Ree I_\gamma$. The brane
loses its exponential suppression when the work pays for the cost,
\begin{equation}
 \frac{W_\gamma}{U_\gamma}\ge1 .
 \label{eq:test}
\end{equation}
This is the criterion we use throughout. The first $\tau$ at which it
is met is the \emph{suppression wall} of $\gamma$, or simply its wall.
At a wall the real part of a probe action vanishes while the cost stays
positive, so no brane becomes tensionless and the wall is not one of
marginal stability. A brane is \emph{actionless} when $I_\gamma$ itself vanishes; for a
real action the two coincide.

Each elementary instanton couples to a single axion, which lowers the action of one orientation and raises that of the other. For the favored orientation, the NS5 has $U = 2\pi s$ and  $W=2\pi\,\Imm a$. The
string on a curve of unit degree has $U=2\pi v$ and $W=2\pi|\Imm b|$. In the R--R regime the same two objects are the bare D5 and the D1.  On the geodesic~\eqref{eq:trajectory} each keeps
\begin{equation}
 \frac{I}{U}=1-\sin u_i ,\qquad i=s,v,
 \label{eq:f1-ns5-actions}
\end{equation}
of its cost, so $W/U=\sin u_i$ reaches one only where its own saxion
diverges, which never happens on a regular wormhole. On the equal-angle wormhole the surviving fraction is smallest at the
end, where it is $1-\sin(\pi\sqrt3/4)\approx0.02$. It multiplies a cost that has
grown along the path, and the actions left at the end are about
$0.66\,v_0$ for the string and $0.66\,s_0$ for the fivebrane. Both are
large when Eq.~\eqref{eq:t} holds, so for a fixed regular division both
instantons remain exponentially suppressed all the way to infinity. The relative
sign decides which orientation of the string is favored, and nothing else
about these two instantons depends on it.

Both are examples of a coupling linear in the charges,
$W_\gamma=2\pi(q_a\Imm a+q_b\Imm b)$. Measuring each axion in units of
its own saxion then separates the brane from the wormhole,
\begin{equation}
 \frac{W_\gamma}{U_\gamma}=\bz_\gamma\cdot\bd,\qquad
 \bz_\gamma=\frac{2\pi}{U_\gamma}\,(q_as,\;q_bv),\qquad
 \bd=\Bigl(\frac{\Imm a}{s},\frac{\Imm b}{v}\Bigr)
    =(\sin u_s,\;\epsilon\sin u_v).
 \label{eq:zY}
\end{equation}
The vector $\bz_\gamma$ is the charge per unit cost of the brane, and
$\bd$ is the local displacement of the axions at its position. The test~\eqref{eq:test}
becomes $\bz_\gamma\cdot\bd\ge1$, the inequality behind the convex-hull
form of the axion weak gravity conjecture~\cite{CheungRemmen2014,Rudelius2015,Brown2015}. The imaginary distance
bound asks for an instanton that satisfies it at the displacement
reached at infinity~\cite{Maldacena2026,Licciardello2026}. 
With fixed charges the same inequality decides whether inserting the brane lowers the action of the wormhole~\cite{Lin2026}.\footnote{Distances
in Ref.~\cite{Lin2026} are $1/\sqrt2$ times ours, so a hyperbolic plane
of radius $R$ has their $\beta=2\sqrt2/R$. Their charge-to-action vector
and wormhole vector are $z^{\rm LS}_i=(\sqrt2/R_i)\,z_i$ and
$Y_i=(R_i/\sqrt2)\,d_i$, with the factors of $2\pi$ absorbed in the
normalization of their axions, so that
$\bz^{\rm LS}\cdot\mathbf Y=\bz\cdot\bd$.} We apply it at every point of
the path, and this is what locates a wall.

The elementary instantons have the unit vectors $(\pm1,0)$ and
$(0,\pm1)$. We call D5 and D1 the orientations with $\bz = (1, 0)$ and $(0, 1)$, which the same-sign wormhole favors together; their reversals are the anti-D5 and the anti-D1. None of them reaches a wall, because each component of $\bd$ stays below one. Its length, however, exceeds one: at the end of
the equal-angle wormhole $|\bd|=\sqrt2\sin(\pi\sqrt3/4)>1$. A brane
whose $\bz$ points along $\bd$ and is longer than $1/|\bd|$ would
therefore have a wall. Whether such branes exist is decided by
two properties. The work is additive: when the coupling is linear, a
brane with charges $\gamma_1+\gamma_2$ receives the sum of the work on
$\gamma_1$ and on $\gamma_2$. The cost is not. At large volume the cost
of a brane is at least a fixed multiple of $|Z_\gamma|$, the modulus of
its central charge, which is linear in $\gamma$; a brane that attains
this bound is \emph{BPS}. A BPS brane of charge $\gamma_1+\gamma_2$,
when it exists, therefore costs at most $U_{\gamma_1}+U_{\gamma_2}$,
with equality only when the two central charges have the same phase,
that is, when $\gamma_1$ and $\gamma_2$ are \emph{mutually
supersymmetric}. Throughout, actions are computed at real values of the fields, where
they are analytic in the axions, and are then evaluated at the imaginary
axions of the wormhole. For a brane coupled linearly to the axions this
affects only the work, so these statements about the cost apply as they
stand. A collection of separately wrapped elementary
instantons adds both work and cost, so its $W/U$ is a cost-weighted
average of $\sin u_s$ and $\sin u_v$ and never reaches one. Among branes whose coupling to the axions is linear,\footnote{The D5 of Sec.~\ref{sec:orig-d5} is not covered: it carries
neither axion charge, and at its wall the action is purely imaginary
rather than zero. Nor is the D5--NS5 bound state of
Sec.~\ref{sec:orientations}, whose coupling to the internal axion is
cubic.} a brane with a wall
must therefore carry both charges and cost strictly less than its
constituents apart: it must be a genuine bound state.

\section{Actionless fivebranes on the two-axion wormhole}
\label{sec:fivebranes}

Section~\ref{sec:criterion} left a definite task. Only a
fivebrane wrapping $X$ couples to the external axion $a$, so we need a
fivebrane that carries string charge.

A fivebrane carries string charge through the field on its worldvolume.
This is the gauge-invariant combination $\mathcal F=F-B$ of the brane's
own gauge field $F$ and the two-form $B$ that it feels: the NS--NS
two-form for a D5, and the RR two-form for an NS5. The internal axion
$b$ can act on the brane in two ways. It can multiply the string
charge that a real magnetic flux $F$ induces, as a theta angle
multiplies a fixed charge; the field is then real. Or $b$ can be part
of the two-form $B$ itself. On the wormhole $b$ is imaginary, so the
field it creates is imaginary, or complex when real flux is also
present. We take the three cases in turn: a real
field (Secs.~\ref{sec:dual-d5} and~\ref{sec:reversal}), an imaginary
field (Sec.~\ref{sec:orientations}) and a complex field
(Sec.~\ref{sec:orig-d5}).

\subsection{A real worldvolume field}\label{sec:dual-d5}

On a threefold, the counterpart of the diagonal string of the
Introduction is a fivebrane with strings dissolved in it as magnetic
flux. This is easiest to analyze in the R--R regime, where the fivebrane that
couples to $a$ is a D5, the string that couples to $b$ is a D1, and
both are elementary D-branes.\footnote{The D1 of the R--R regime is the
fundamental string of the original variables, so on a curve of unit
degree it costs $2\pi v$, not $2\pi t$.} Let the D5 carry $k$ units of
magnetic flux through each curve of unit degree, $F=kH$ with $k$ an
integer. Its action follows from three standard properties of a
D-brane with flux.

\emph{Cost.} The Dirac--Born--Infeld (DBI) action is the integral of
$\sqrt{\det(g+F)}$ over the brane. At each point the tangent space of
$X$ is $\mathbb C^3$, and the K\"ahler form $\tilde J=tH$ of the S-dual
description [Eq.~\eqref{eq:tilde}] is the sum of the area forms of three
orthogonal complex planes. Since $F=(k/t)\tilde J$, the field has the
same strength $x=|k|/t$ relative to the area in each of them, and each
plane multiplies the action by $\sqrt{1+x^2}$. The bare D5 has action
$2\pi\tilde V/\tilde g_s=2\pi s$, with $\tilde V=\kappa t^3/6$, so
\begin{equation}
 U_k=2\pi s\,(1+x^2)^{3/2}.
 \label{eq:cost}
\end{equation}

\emph{Charge.} Through its Wess--Zumino coupling, flux in two of the three planes induces
string charge along the third, in total $\int_XH\wedge F^2/2=\kappa
k^2/2$ anti-D1s on curves of unit degree. The charge is quadratic in
the flux, so it is anti-D1 charge for either sign of $k$. That it is anti-D1 and
not D1 charge is read off from the central charge~\eqref{eq:Zk} below,
and it fixes our sign convention: with $\tilde C_2=B_2$ as in
Eq.~\eqref{eq:tilde}, the brane couples to the two axions through
$2\pi i\int_X\bigl(\tilde C_6-\tilde C_2\wedge F^2/2\bigr)
=2\pi i\,(a-\tfrac12\kappa k^2\,b)$. Equation~\eqref{eq:reverse} and
Appendix~\ref{app:flux-bounds} use the same convention. The flux also
induces D3 and D($-1$) charge, but $\tilde C_4$ and $\tilde C_0$ vanish
on the wormhole and these charges play no role in the work.

\emph{Work.} The D5 charge and the anti-D1 charge are the two
elementary instantons of Sec.~\ref{sec:criterion}, and the axions do
the same work on them whether or not they are bound. The D5 charge
receives $2\pi s\,d_a$. Wrapped separately, the anti-D1s would cost
$2\pi v\cdot\kappa k^2/2=2\pi s\cdot3x^2$, where the factor of three
counts the pairs of planes, and they receive $-2\pi s\cdot3x^2d_b$. So
\begin{equation}
 W_k=2\pi s\,(d_a-3x^2d_b),
 \label{eq:corework}
\end{equation}
and the action of the brane at any point of either wormhole is
\begin{equation}
 I_k=2\pi s\,\bigl[(1+x^2)^{3/2}-d_a+3x^2d_b\bigr],
 \label{eq:split}
\end{equation}
with $x=|k|/t(\tau)$ and $\bd=(d_a,d_b)$ the displacement of
Eq.~\eqref{eq:zY}. At $\bd=0$ it is the cost. On the opposite-sign
wormhole $d_b=-\sin u_v$, so the axions do positive work on both
charges.

\begin{figure}[tbp]
\centering
\begin{tikzpicture}[x=3.55cm,y=3.0cm,font=\footnotesize]
 \node[anchor=west,font=\small] at (-.2,1.30) {(a) Action along the wormhole};
 \draw[->] (0,-.6) -- (0,1.10) node[right] {$I/U$};
 \draw[->] (0,-.6) -- (2.92,-.6) node[above] {$\tau$};
 \draw[black!45,thin] (0,0) -- (2.80,0);
 \foreach \y/\lab in {-.5/{-0.5},0/0,.5/{0.5},1/1} \draw (0,\y) -- ++(-.025,0) node[left] {$\lab$};
 \foreach \xx/\lab in {0/0,1.57080/{\pi/2},2.09440/{2\pi/3},2.72070/{\tau_{\rm end}}}
   \draw (\xx,-.6) -- ++(0,-.04) node[below] {$\lab$};
 \draw[black!35,densely dotted] (1.57080,-.6) -- (1.57080,0);
 \draw[black!35,densely dotted] (2.09440,-.6) -- (2.09440,0);
 \draw[black!45,densely dotted] (2.72070,-.6) -- (2.72070,1.0);
 \draw[black!70,thick,dashed] plot[domain=0:2.72070,samples=80]
   (\x,{1-sin(deg(.5*\x))});
 \draw[orange!80!black,very thick,densely dotted] plot[domain=0:2.72070,samples=80]
   (\x,{2^1.5*cos(deg(.5*\x))^1.5*cos(deg(.75*\x))/((1+cos(deg(.5*\x))^2)^1.5)});
 \draw[blue!65!black,very thick,dash dot] plot[domain=0:2.72070,samples=80]
   (\x,{2*cos(deg(.5*\x))*(sqrt(2)*cos(deg(.5*\x))^2-sin(deg(.5*\x)))/((1+cos(deg(.5*\x))^2)^1.5)});
 \draw[teal!75!black,very thick] plot[domain=0:2.72070,samples=80]
   (\x,{1-sqrt(2)*sin(deg(.5*\x))});
 \fill[black] (1.57080,0) circle[radius=1.8pt] (2.09440,0) circle[radius=1.8pt];
 \node[anchor=west,orange!75!black] at (1.30,.66) {D5 of the NS--NS regime (real part)};
 \node[anchor=west,black!80] at (2.06,.25) {elementary};
 \node[anchor=east,teal!65!black] at (.66,.36) {fluxed D5};
 \node[anchor=east,blue!65!black] at (2.05,-.47) {D5--NS5 bound state};
\end{tikzpicture}

\vspace{5mm}
\begin{tikzpicture}[x=2.3cm,y=2.3cm,font=\footnotesize,baseline=(current bounding box.north)]
 \node[anchor=west,font=\small] at (-.35,.62) {(b) Opposite-sign wormhole};
 \draw[->] (-.3,0) -- (1.7,0) node[right] {$z_a$};
 \draw[->] (0,-1.5) -- (0,.45) node[right] {$z_b$};
 \draw[black!70,thick,dashed] (1,0) -- (0,-1);
 \draw[black!55,thin] (0,-1.41421) -- (1.41421,0);
  \draw[teal!75!black,very thick] plot[domain=0:90,samples=180,variable=\th]
   ({cos(\th)^3},{-3*sin(\th)^2*cos(\th)});
 \fill[teal!75!black] (.35355,-1.06066) circle[radius=1.8pt];
 \fill[black!80] (1,0) circle[radius=1.5pt] (0,-1) circle[radius=1.5pt];
 \draw[->,thick,teal!70!black] (0,0) -- (.24,-.24);
 \node[teal!70!black] at (.34,-.17) {$\bd$};
 \node[anchor=south west] at (1,0) {D5};
 \node[anchor=east] at (-.03,-1) {$\overline{\rm D1}$};
 \node[anchor=west,teal!65!black] at (.42,-1.14) {$|k|=t_0$};
 \node[anchor=west,teal!65!black,align=left] at (.06,-.50) {fluxed\\D5};
 \node[anchor=west,black!70,align=left] at (1.0,-.55) {$\bz\cdot\bd=1$\\ at $\tau=\pi/2$};
\end{tikzpicture}
\hspace{10mm}
\begin{tikzpicture}[x=2.3cm,y=2.3cm,font=\footnotesize,baseline=(current bounding box.north)]
 \node[anchor=west,font=\small] at (-.35,1.62) {(c) Same-sign wormhole};
 \fill[black!10] (0,1) -- (1,0) -- (1.65,0) -- (1.65,1.4) -- (0,1.4) -- cycle;
 \draw[->] (-.3,0) -- (1.7,0) node[right] {$z_a$};
 \draw[->] (0,-.45) -- (0,1.45) node[right] {$z_b$};
 \draw[black!70,thick,dashed] (0,1) -- (1,0);
  \draw[black!55,thin] (1.41421,0) -- (0,1.41421);
 \draw[blue!65!black,very thick,dash dot] (.3536,.3536) -- (.7918,.7918);
 \fill[blue!65!black] (.70711,.70711) circle[radius=1.8pt];
 \fill[black!80] (1,0) circle[radius=1.5pt] (0,1) circle[radius=1.5pt];
 \draw[->,thick,blue!65!black] (0,0) -- (.22,.22);
 \node[blue!65!black] at (.30,.12) {$\bd$};
 \node[anchor=north] at (1,-.02) {D5};
 \node[anchor=east] at (-.03,1) {D1};
 \node[anchor=west,blue!65!black,align=left] at (.86,.84) {D5--NS5\\(effective)};
 \node[anchor=west,black!70,align=left] at (1.22,.40) {$\bz\cdot\bd=1$\\ at $\tau=\pi/2$};
 \node[black!60] at (1.30,1.22) {no real flux};
\end{tikzpicture}
\caption{The equal-angle wormhole at leading order at large volume.
(a) The action of each brane divided by its cost at the same point,
against the distance from the throat. The fluxed D5 is on the
opposite-sign wormhole and the D5--NS5 bound state on the same-sign one;
both vanish at $\pi/2$. For the D5 of the NS--NS regime, on either
wormhole, the curve is the real part, which vanishes at $2\pi/3$. The
elementary instantons stay positive to the end.
(b) The opposite-sign wormhole in the plane of the charge per unit cost
$\bz$. Collections of D5s and anti-D1s lie on the dashed segment, which
is a level line of $\bz\cdot\bd$. The fluxed D5 moves along the solid
curve as $|k|/t$ grows, and at $|k|=t_0$ it touches the level
$\bz\cdot\bd=1$ at $\tau=\pi/2$.
(c) The same-sign wormhole. No D-brane with smooth Abelian flux lies in
the shaded region beyond the dashed line from D1 to D5. The effective
vector of the D5--NS5 bound state, followed along the wormhole, sweeps
the dash-dotted segment and crosses the level $\bz\cdot\bd=1$ at
$\tau=\pi/2$.}
\label{fig:flux}
\end{figure}
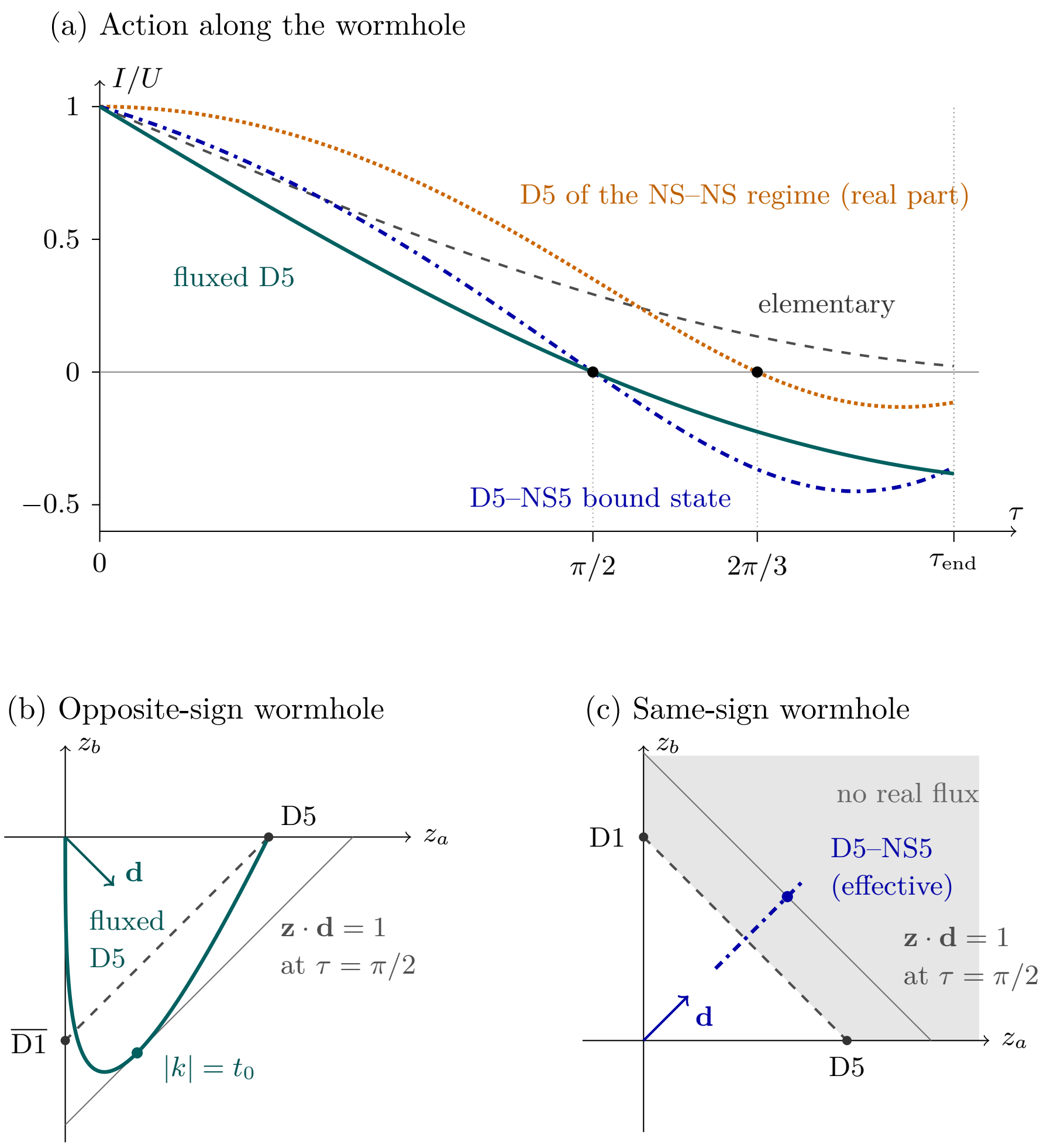

Is this enough to produce a wall? Take the opposite-sign wormhole at
equal angles, for which $t=t_0$ is constant along the path and
$\bd=(\sin\frac\tau2,-\sin\frac\tau2)$. Then
\begin{equation}
 \frac{I_k}{U_k}=1-\frac{1+3x^2}{(1+x^2)^{3/2}}\,\sin\frac\tau2 .
 \label{eq:main}
\end{equation}
Flux helps when the coefficient of the sine exceeds one. It rises from
one at $x=0$ to a maximum of $\sqrt2$ at $x=1$ and then falls. With one
unit of flux per unit area the work is four times that on a bare D5,
while the cost is only $2\sqrt2$ times larger. The brane with
$|k|=t_0$ therefore has\footnote{The wall is exactly at $\pi/2$ when
$t_0$ is an integer. Otherwise the nearest integer gives a wall later
than $\pi/2$ by an amount of order $t_0^{-2}$, since the maximum at
$x=1$ is quadratic.}
\begin{equation}
 \frac{I_k}{U_k}=1-\sqrt2\,\sin\frac\tau2,\qquad
 \tau_*=\frac\pi2<\tau_{\rm end} .
 \label{eq:threshold}
\end{equation}
Its action vanishes well before the end of the wormhole, at a point
where each elementary instanton still has a fraction $1-1/\sqrt2$ of
its cost [Fig.~\ref{fig:flux}(a)]. The intersection number $\kappa$
has dropped out, and the optimal flux has a simple meaning. At
$|k|=t_0$ the string charge per unit fivebrane charge, $\kappa k^2/2$,
equals the ratio $|n_b/n_a|=\kappa t_0^2/2$ of the conserved fluxes of
the equal-angle wormhole, Eq.~\eqref{eq:fluxratio}. At equal angles the
brane that loses its action first is the one whose charges point along
those of the background.

Why does binding help so much? Because the D5 and the anti-D1 are not
mutually supersymmetric. In the R--R regime the cost of a BPS D-brane
is $2\pi|Z_\gamma|/\tilde g_s$, with $Z_\gamma=\int_Xe^{-i\tilde
J}\gamma$. For the fluxed D5, $\gamma=e^{kH}$ and
\begin{equation}
 Z_k=\frac\kappa6\,(k-it)^3
 =i\tilde V\,\bigl[(1-3x^2)+i\,(3x-x^3)\bigr]
 \label{eq:Zk}
\end{equation}
for $k>0$. The four terms in the bracket are the D5, string, D3 and
D($-1$) contributions, in that order. The two on which the axions act,
the D5 and string terms, have opposite signs: the D5 and the string
charge it carries have opposite phases. They partly cancel inside the
modulus, $|Z_k|=\tilde V(1+x^2)^{3/2}$. This partial cancellation
explains why, at the optimal flux $x=1$, the bound state has
$|Z_k|=2\sqrt2\,\tilde V$, smaller than the $4\tilde V$ of the same D5
and anti-D1 charges wrapped separately, despite the additional induced
D3 and D($-1$) charges.
This is the sub-additivity anticipated in Sec.~\ref{sec:criterion}.

Figure~\ref{fig:flux}(b) shows the same gain in the plane of the
charge per unit cost $\bz$ of Sec.~\ref{sec:criterion}. Separately
wrapped D5s and anti-D1s add both work and cost, so any collection of
them lies on the straight segment from $(1,0)$ to $(0,-1)$. At equal
angles on the opposite-sign wormhole $\bd$ is proportional to $(1,-1)$,
so this segment is a level line of $\bz\cdot\bd$, and no such
collection does better than a single elementary instanton. By
Eq.~\eqref{eq:corework} the flux moves the D5 along the curve
\begin{equation}
 \bz_k=\frac{(1,\,-3x^2)}{(1+x^2)^{3/2}},
 \label{eq:zk}
\end{equation}
which starts at the bare D5 and bulges beyond the segment, furthest in
the direction of $\bd$ at $x=1$. There it touches the level
$\bz\cdot\bd=1$ of the wall at $\tau=\pi/2$.

It remains to check that such a brane exists. Its charge is the class
of the line bundle $\mathcal O_X(k)$ in $K^0(X)$, with cohomological
representative~\cite{Alexandrov2010}
\begin{equation}
 \gamma_k=e^{kH}\bigl(1+c_2(X)/24\bigr),
 \label{eq:mukai}
\end{equation}
in which the curvature term is negligible at large volume, and the
field $F=(k/t)\tilde J$ minimizes the DBI action at real
moduli~\cite{Marino1999,Alexandrov2021}. The integer $k$ belongs to the
brane and is independent of the fluxes $(n_a,n_b)$ of the wormhole. No
other single D-brane with smooth Abelian flux does better. A bound on
the work in each tangent plane gives $W/U\le\sqrt2\sin(\tau/2)$ for any
magnetic field on the D5 and $W/U\le\sin(\tau/2)$ for a D3, D1 or
D($-1$) (Appendix~\ref{app:flux-bounds}). Whether a stable brane of
higher rank has an earlier wall is open, and an earlier wall would only
strengthen the conclusion.

\subsection{Why real flux fails on the same-sign wormhole}\label{sec:reversal}

Now take the same-sign wormhole, obtained by reversing the internal axion at the
same throat. This acts only on the wormhole, as the reflection
\begin{equation}
 R:\;(d_a,d_b)\mapsto(d_a,-d_b),
 \label{eq:reflection}
\end{equation}
and the elementary instantons cannot tell the difference: the favored
string is now the D1 instead of the anti-D1, with the same cost and the
same work. For the fluxed D5 the term $3x^2d_b$ in Eq.~\eqref{eq:split}
changes sign, and at equal angles
\begin{equation}
 \frac{W_k}{U_k}=\frac{1-3x^2}{(1+x^2)^{3/2}}\,\sin\frac\tau2
 \le\sin\frac\tau2 .
 \label{eq:reversed-flux}
\end{equation}
Flux now makes matters worse, because the anti-D1 charge it induces is
disfavored. In the central charge~\eqref{eq:Zk}, $1-3x^2$ is the real
part of the bracket, whose modulus is the cost, and it can never exceed
that modulus.

The same holds for a BPS D-brane of any charge. The favored charges on
the same-sign wormhole are those of the D5 and the D1, which are
mutually supersymmetric: their central charges are aligned, their costs
add, and nothing is gained by binding them. In formulas, a BPS D-brane
of charge $\gamma$ costs $2\pi|Z_\gamma|/\tilde g_s$, and its D5 and D1
charges, the components $\gamma_0$ and $\gamma_4$ of degree zero and
four, enter the imaginary part
\begin{equation}
 \Imm Z_\gamma=\tilde V\gamma_0-\int_X\tilde J\wedge\gamma_4 .
 \label{eq:reverse}
\end{equation}
At equal angles the work of the same-sign wormhole is $\sin(\tau/2)$
times exactly this combination, so
\begin{equation}
 \frac{|W_\gamma|}{U_\gamma}
 =\frac{|\Imm Z_\gamma|}{|Z_\gamma|}\,\sin\frac\tau2
 \le\sin\frac\tau2 .
 \label{eq:dbrane-bound}
\end{equation}

The bound does not rely on supersymmetry of the brane. It holds point
by point on the worldvolume, for any smooth Abelian magnetic field
(Appendix~\ref{app:flux-bounds}). In each complex tangent plane a field
of strength $\tan\alpha_i$ relative to the area turns the phase of the
local central charge by $\alpha_i$, and the ratio of the local work to
the local cost is $|\cos(\alpha_1+\alpha_2+\alpha_3)|\sin(\tau/2)$. At
equal angles, then, no D-brane with smooth Abelian flux has a wall on
the same-sign wormhole: none lies beyond the dashed line from D1 to D5
in Fig.~\ref{fig:flux}(c).

The bound holds because the angles $\alpha_i$ are real, so the cosine
cannot exceed one. For an imaginary field strength the angles would be
imaginary, and the cosine would become a hyperbolic cosine, which does
exceed one.

\subsection{An imaginary worldvolume field}\label{sec:orientations}

The continued background supplies an
imaginary field, in the form of the two-form itself. In the R--R regime
the imaginary two-form is $\tilde C_2$, while $\tilde B_2=0$, so it is
an NS5 that sees it in its worldvolume field. An NS5 alone is not
enough, because it does not couple to $\tilde C_6$ and the external
axion does no work on it. The brane we need has both components. Let it
consist of $p$ D5s and $q$ NS5s, with $p,q>0$ coprime and $F=0$.

Its action follows from a change of description. Type IIB has an
$SL(2,\mathbb Z)$ duality under which a coprime $(p,q)$ fivebrane is
the image of a single D5. In the description where it is a single D5,
which we mark with hats, the background is~\cite{Alexandrov2010}
\begin{equation}
 \hat g_s=\tilde g_s\,\Delta^2,\qquad
 \hat J=\Delta\,\tilde J,\qquad
 \hat B_2=q\,\tilde C_2,\qquad
 \Delta^2=p^2+\frac{q^2}{\tilde g_s^2} .
 \label{eq:hat-frame}
\end{equation}
This description is not weakly coupled, and we use it only to organize
the calculation. The brane is the familiar non-threshold bound state of
D5 and NS5 branes~\cite{LuRoy1998,Aharony1997}, a BPS fivebrane instanton of the compactification at
real moduli. Its action there is fixed by its charges and by duality,
as is the tension of a $(p,q)$ string, and we continue it to the
wormhole background.

The first two relations in Eq.~\eqref{eq:hat-frame} give the cost. With
$\hat J=\hat t\,H$, $\hat t=\Delta t$ and $\hat V=\kappa\hat t^{\,3}/6=\Delta^3\tilde V$, the action 
$2\pi\hat V/\hat g_s=2\pi s\Delta$ of the bare brane is the
hypotenuse formed by $p$ D5s, of action $2\pi s$ each, and $q$ NS5s, of
action $2\pi s/\tilde g_s$ each. The third relation is the important
one. The RR two-form of the wormhole appears as a $B$-field, and the D5
sees it as a worldvolume field $\mathcal F=-\hat B_2=-qbH$, which is
imaginary. The DBI action is that of Sec.~\ref{sec:dual-d5} with $x^2$
replaced by
\begin{equation}
 \frac{\mathcal F^2}{\hat t^{\,2}}
 =-\frac{q^2(\Imm b)^2}{\Delta^2t^2}=-\nu\,d_b^2,\qquad
 \nu\equiv\frac{q^2}{\tilde g_s^2\Delta^2},
 \label{eq:nu}
\end{equation}
where we used $\Imm b=v\,d_b$ and $v=t/\tilde g_s$. The parameter
$\nu$ is the share of the NS5s in the squared tension. Where a real
field raises the DBI action, an imaginary one lowers it.

The Wess--Zumino coupling has two parts, as before. The $p$ units of D5 charge couple to the external axion through
$2\pi i\,p\int_X\tilde C_6=2\pi i\,p\,a\to-2\pi s\,p\,d_a$, which is $p$ times
the work on a bare D5. The second part is new, and it comes from the duality itself. The
$SL(2,\mathbb Z)$ matrix that turns the bound state into a single D5 has
bottom row $(q,p)$. Acting on the background, in which the RR scalar
vanishes, it does more than rescale the coupling as in
Eq.~\eqref{eq:hat-frame}: it rotates part of the dilaton into an RR
scalar,
\begin{equation}
 \hat C_0=-\frac{p}{q\,\Delta^2}+\frac{n}{q},
 \label{eq:hatC0}
\end{equation}
where the integer $n$, the upper-left entry of the matrix, is defined
only modulo $q$. A D5 whose worldvolume field is $\mathcal F$ carries
D-instanton charge $\int_X\mathcal F^3/6$, and this charge couples to
the RR scalar. In Sec.~\ref{sec:dual-d5} the same coupling was present
but the scalar vanished; here the duality has made it nonzero. With
$\mathcal F=-qbH$, the first term of Eq.~\eqref{eq:hatC0} gives a term
in the action,
\begin{equation}
 -2\pi i\,\frac{\kappa}{6}\,\frac{p\,q^2}{\Delta^2}\,b^3
 \;\longrightarrow\;
 -2\pi s\,p\,\nu\,d_b^3 ,
 \label{eq:cubic}
\end{equation}
where the arrow evaluates it on the wormhole, with $b=i\,\Imm b$,
$\Imm b=v\,d_b$ and $\kappa v^3/6=s/\tilde g_s^2$; its overall sign is
fixed by the convention of Sec.~\ref{sec:dual-d5}. Nothing else
survives. The terms of $\int\hat C\wedge e^{\mathcal F}$ that are linear
and quadratic in $\mathcal F$ cancel each other, because the duality
gives $\hat C_4=\tfrac12\hat B_2\wedge\hat C_2$ when $\tilde C_4$ and
$\tilde B_2$ vanish. The term $n/q$, which depends on an arbitrary
choice, cancels against a term cubic in $\tilde C_2$ in the six-form
potential, as it must. The coupling~\eqref{eq:cubic} is therefore cubic
in the internal axion and proportional to the number of D5s. It is
induced by the background and is not a quantized charge of the brane.
It has the sign of $d_b$, so on the same-sign wormhole it adds to the
first part. The same term appears in the phase of the fivebrane
amplitude~\cite[Eqs.~(4.29), (4.40)--(4.41)]{Alexandrov2010}.
Altogether
\begin{equation}
 I_{p,q}=2\pi s\,\Bigl[\Delta\,(1-\nu\,d_b^2)^{3/2}
 -p\,\bigl(d_a+\nu\,d_b^3\bigr)\Bigr].
 \label{eq:mixed-general}
\end{equation}
This has the structure of Eq.~\eqref{eq:split} with the real field
replaced by an imaginary one, and at $\bd=0$ it is the cost
$U_{p,q}=2\pi s\Delta$. The imaginary field does two things, both
contained in the work $W=U-I$: it lowers the DBI term, and it induces a
coupling of the favored sign.

The brane is most effective when its two components have equal tension
at the throat, $p/q=1/\tilde g_{s,0}=g_{s,0}$, so that $\nu=1/2$ there.
At equal angles $\tilde g_s=\tilde g_{s,0}\cos(\tau/2)$, and with
$u=\tau/2$ Eq.~\eqref{eq:mixed-general} becomes
\begin{equation}
 \frac{I_{p,q}}{U_{p,q}}
 =\frac{2\cos u\,\bigl(\sqrt2\cos^2u-\sin u\bigr)}{(1+\cos^2u)^{3/2}} .
 \label{eq:mixed-real}
\end{equation}
It vanishes where $\sin u=\sqrt2\cos^2u$, that is, at $u=\pi/4$, so
\begin{equation}
 \tau_*=\frac\pi2
\end{equation}
once more. The action is a different function of $\tau$ from that of
the fluxed D5, but its zero is at the same place
[Fig.~\ref{fig:flux}(a)]. For other ratios the zero is later, by an
amount quadratic in the mismatch. The condition asks for many branes. In
the R--R regime $g_{s,0}\gg1$, so $p\approx g_{s,0}\,q$ is large (in the
NS--NS regime it is $q$ that is large), and the cost $2\pi s\Delta$
grows with it. The brane is a probe only if the fluxes of the wormhole
are larger still, $n_a\gg p$. For irrational $g_{s,0}$ the distance
$\pi/2$ is an infimum, approached by coprime pairs of ever larger
charges, under the same condition.

The fivebrane charges of this brane are fixed, but its dependence on
the two axions is not a linear coupling with a constant charge vector,
so it has no fixed point in the plane of $\bz$. What it has is an
effective vector $(W/U)\,\bd/|\bd|^2$, the charge per unit cost that a
linearly coupled brane would need in order to receive the same work.
Followed along the wormhole, this vector sweeps the dash-dotted segment
of Fig.~\ref{fig:flux}(c). At $\tau=\pi/2$ it crosses the level
$\bz\cdot\bd=1$, beyond the line that no D-brane with smooth flux can
cross.

\subsection{A complex worldvolume field}\label{sec:orig-d5}

In the NS--NS regime the two branes above are an NS5 with $k$ units of
worldvolume flux and a bound state with wrappings $(q,-p)$. Both are
solitons there, and their actions are the ones already found. The
elementary D-brane of this regime, the D5, gives the third case. All RR
potentials vanish, so the axions do no work through its charges. The
internal axion enters its worldvolume field instead,
$\mathcal F=F-B_{\rm int}=(k_{\rm D}-b)H$, which is complex once $b$ is
imaginary. Each complex tangent plane contributes the same factor
$\sqrt\Xi$ to the DBI action, with the square root continued from real
moduli, and~\cite{Alexandrov2021}
\begin{equation}
 I_{k_{\rm D}}=\frac{2\pi\kappa}{6g_s}\bigl(\sqrt\Xi\bigr)^3,\qquad
 \Xi=v^2+(k_{\rm D}-b)^2 .
 \label{eq:original-dbi}
\end{equation}
Without flux $\Xi=v^2-(\Imm b)^2=v_0^2$ is real, and the action stays
real and positive. Flux makes $\Xi$ complex,
$\Xi=v_0^2+k_{\rm D}^2-2i\epsilon k_{\rm D}v_0\tan u_v$, and its phase
grows fastest for $|k_{\rm D}|=v_0$, where
$\Xi=2v_0^2\sec u_v\,e^{\mp iu_v}$. Each square root then rotates by
$u_v/2$, and
\begin{equation}
 \Ree I_{k_{\rm D}}=|I_{k_{\rm D}}|\cos\frac{3u_v}{2},\qquad
 \tau_*^{\rm D5}=\frac{\pi}{\sqrt3\,p_v}.
 \label{eq:original-wall}
\end{equation}
This is $2\pi/3$ at equal angles [Fig.~\ref{fig:flux}(a)], and since
the action depends only on $u_v$, it holds for every division of the
excursion. It precedes the end on every regular wormhole, since
$p_v>\sqrt{2/3}$ there. Reversing the internal axion conjugates the
action, so both signs give the same wall.

Nothing here subtracts from the cost: the complex field only turns the
phase of the action. So the wall is of a different kind. The action does
not vanish there. It is purely imaginary, and the weight $e^{-I}$ is a
pure phase: the brane loses its exponential suppression without
becoming actionless. The value of this brane is that it shows, by a
D-brane calculation in the NS--NS regime, that the wormhole does not
reach its end with every instanton suppressed.

\subsection{Every division of the excursion}\label{sec:root-wall}

At equal angles two branes with different actions have their walls at
the same distance. Is this special to equal angles? Let the two
hyperbolic planes share the excursion unequally. Then $t$ changes along
the path, and a fluxed D5 of fixed $k$ no longer has fixed $x$.
Minimizing Eq.~\eqref{eq:split} over $x$ at each point bounds the
action of every fluxed D5 on the opposite-sign wormhole from below. For
$u_v>\pi/6$ the minimum is at $x^2=4\sin^2u_v-1$, and the triple-angle
identity gives
\begin{align}
 \min_{x}\frac{I_k}{2\pi s}
 &=\sin3u_v-\sin u_s \nonumber\\
 &=2\cos\frac{u_s+3u_v}{2}\,\sin\frac{3u_v-u_s}{2}.
 \label{eq:root-envelope}
\end{align}
For smaller $u_v$ the bare D5 is optimal and its action is positive.

On the same-sign wormhole both terms of Eq.~\eqref{eq:mixed-general}
are positive up to the wall, where $1-\nu d_b^2$ is still positive, so
the action has the sign of the difference of their
squares. In units of $(2\pi s\Delta)^2$, completing a square and using
the triple-angle identity once more, this difference is
\begin{equation}
 \bigl[(1-\nu)\cos u_s-\nu\cos^3u_v\bigr]^2
 +4\nu(1-\nu)\cos\frac{u_s+3u_v}{2}\,\cos^3\frac{u_s-u_v}{2}.
 \label{eq:mixed-root-gap}
\end{equation}
The square cannot be negative, so no ratio $p/q$ reaches a wall before
the product of cosines does, and the ratio that cancels the square
reaches it there. In Eqs.~\eqref{eq:root-envelope} and~\eqref{eq:mixed-root-gap} the
last factor is positive on a
regular wormhole, and the first vanishes at
\begin{equation}
 \frac{u_s+3u_v}{2}=\frac\pi2,\qquad
 \tau_*=\frac{\pi}{p_s+\sqrt3\,p_v}.
 \label{eq:root-distance}
\end{equation}

A single brane attains this bound under the same conditions as before,
$|k|=t_0$ or $p/q=g_{s,0}$, for every division. For the flux, the
geodesic gives $t^2/t_0^2=\cos u_v/\cos u_s$, and at the wall
$\cos u_s=(4\sin^2u_v-1)\cos u_v$, so the optimal
$k^2=t^2(4\sin^2u_v-1)$ equals $t_0^2$. Before the wall the action of
this brane lies above the envelope~\eqref{eq:root-envelope}, and the two
meet at the zero. For the bound state,
$(1-\nu)/\nu=(p\tilde g_s/q)^2$ and
$\tilde g_s^2/\tilde g_{s,0}^2=\cos^3u_v/\cos u_s$, so $p/q=g_{s,0}$
cancels the square in Eq.~\eqref{eq:mixed-root-gap} along the entire
path.

Over the regular range $0<p_s<1/\sqrt3$ the sum $p_s+\sqrt3\,p_v$ lies
between $\sqrt3$ and $2$, which gives Eq.~\eqref{eq:headline}. The wall
is never further than two thirds of the way to the end, and it is
earliest, at $\pi/2$, for equal angles. The D5 of the NS--NS regime,
with its wall at $\pi/(\sqrt3\,p_v)$, always comes
later.\footnote{\label{fn:reversed}Reversing the orientation of either
fivebrane, on the wormhole of the other sign, gives a second pair of
walls at $\tau=\pi/(\sqrt3\,p_v-p_s)$: the anti-D5 with flux on the
same-sign wormhole and the bound state $(-p,q)$ on the opposite-sign
one. They lie before the end only when $p_s<(\sqrt2-1/\sqrt3)/2$, and
always after the walls above.} Table~\ref{tab:branes} lists the branes
with walls found here.

\begin{table}[t]
\centering
\renewcommand{\arraystretch}{1.4}\setlength{\tabcolsep}{3pt}\small
\begin{tabular}{@{}lllll@{}}
\toprule
Brane & Worldvolume field & Wormhole & Wall & Sec. \\
\midrule
D5 with magnetic flux, $|k|=t_0$ & real & opposite-sign &
 $\pi/(p_s+\sqrt3\,p_v)$ & \ref{sec:dual-d5} \\
D5--NS5 bound state, $p/q=g_{s,0}$ & imaginary & same-sign &
 $\pi/(p_s+\sqrt3\,p_v)$ & \ref{sec:orientations} \\
D5 of the NS--NS regime, $|k_{\rm D}|=v_0$ & complex & either &
 $\pi/(\sqrt3\,p_v)$ & \ref{sec:orig-d5} \\
\midrule
anti-D5 with flux, $|k|=t_0$ & real & same-sign &
 $\pi/(\sqrt3\,p_v-p_s)$ & \ref{sec:root-wall} \\
bound state $(-p,q)$, $p/q=g_{s,0}$ & imaginary & opposite-sign &
 $\pi/(\sqrt3\,p_v-p_s)$ & \ref{sec:root-wall} \\
\bottomrule
\end{tabular}
\caption{The branes with walls found here on the two-axion wormhole, at leading
order at large volume. All but the third are described in the R--R
regime, and their actions vanish at the wall. The third is the D5 of the
NS--NS regime, and at its wall the real part of its action vanishes. The
last two walls lie before the end only when
$p_s<(\sqrt2-1/\sqrt3)/2\approx0.42$.}
\label{tab:branes}
\end{table}

These results raise four questions.
\begin{enumerate}
\item[(i)] Why do two branes with different actions have the same wall,
for every division of the excursion?
\item[(ii)] Why does the wall not depend on the relative sign, while the
brane does?
\item[(iii)] Why does the complex field give a wall that is later, and
of a different kind?
\item[(iv)] Why do equal angles give the earliest wall?
\end{enumerate}

\subsection{The geometry of the wall}\label{sec:geometry}

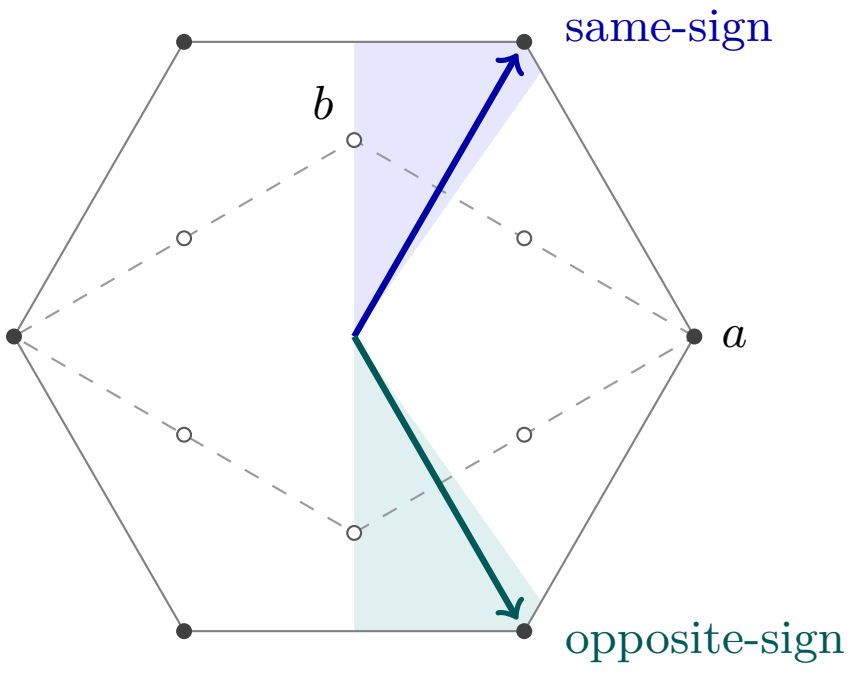
\begin{figure}[tbp]
\centering
\begin{tikzpicture}[x=2.25cm,y=2.25cm,font=\footnotesize]
 \begin{scope}
 \clip (0:1) \foreach \ang in {60,120,180,240,300}{-- (\ang:1)} -- cycle;
 \fill[blue!10] (0,0) -- (54.7356:1.2) arc (54.7356:90:1.2) -- cycle;
 \fill[teal!12] (0,0) -- (-54.7356:1.2) arc (-54.7356:-90:1.2) -- cycle;
 \end{scope}
 \draw[black!50] (0:1) \foreach \ang in {60,120,180,240,300}{-- (\ang:1)} -- cycle;
 \draw[black!40,dashed] (1,0) -- (0,.57735) -- (-1,0) -- (0,-.57735) -- cycle;
 \foreach \ang in {0,60,...,300} \fill[black!75] (\ang:1) circle[radius=1.5pt];
 \foreach \ang in {30,90,...,330} \draw[black!65,fill=white] (\ang:.57735) circle[radius=1.3pt];
 \draw[->,very thick,blue!65!black] (0,0) -- (60:.96);
 \draw[->,very thick,teal!70!black] (0,0) -- (-60:.96);
 \node[right] at (1.02,0) {$a$};
 \node[above left] at (0,.58) {$b$};
 \node[right,blue!65!black] at (.56,.90) {same-sign};
 \node[right,teal!70!black] at (.56,-.90) {opposite-sign};
\end{tikzpicture}
\caption{Roots of the one-modulus scalar space $G_{2(2)}/SO(4)$ in
coordinates orthonormal for Eq.~\eqref{eq:metric}. Filled dots are long
roots and open dots short roots. The dashed rhombus joins the two
elementary directions $a,b$ and their negatives. The solid hexagon is
the convex hull of all the roots. Regular wormholes on the mouth with
$\Imm a>0$ point into the shaded sectors, and the arrows are the two
equal-angle wormholes. In both directions the rhombus is at its
narrowest and the hexagon has a vertex.}
\label{fig:root-angles}
\end{figure}

The two hyperbolic planes of Eq.~\eqref{eq:metric} are a slice of a
larger symmetric space, which also contains the RR axions. For
$h^{1,1}=1$ the tree-level scalar manifold that follows from the cubic
prepotential is locally
$G_{2(2)}/SO(4)$~\cite{Gunaydin2007,Chiodaroli2009}. 
The same holds on the slice of any threefold on which the K\"ahler form
and the two-form potentials point along $H$ and the four-form potential
along $H^2$. There the K\"ahler metric $G_J=-\partial_J^2\log V$ maps $H$ to $H^2$, since
$G_J(J,\cdot\,)=\dd\log V$ for a cubic volume, and the metric on the
slice depends on the intersection numbers only through $\kappa=H^3$. This space has two
saxions and six axions, one for each positive root, and the root of an
axion records how the cost of its instanton scales with the saxions. In
the NS--NS regime a D-brane on a $2j$-cycle costs
$v^j/g_s\propto s^{1/2}v^{j-3/2}$, so the RR axions have weights
$(\tfrac12,j-\tfrac32)$ in the basis $(\log s,\log v)$, for
$j=0,\dots,3$. Our two hyperbolic planes are the slice on which these
four RR axions vanish. In it $a$ is a long root and $b$ an orthogonal
short one, which is the origin of the radii $1$ and $\sqrt3$. The outer
pair $(\tfrac12,\pm\tfrac32)$, of the D($-1$) and the D5, are long roots
(Fig.~\ref{fig:root-angles}). The figure uses coordinates orthonormal
for Eq.~\eqref{eq:metric}. In them the vectors $\bz$ of
Eq.~\eqref{eq:zY} are divided by the radii of their hyperbolic planes,
so that the string has length $1/\sqrt3$ instead of one. We use this
normalization from here on.

Each root $\alpha$ spans a hyperbolic plane with its own saxion
direction. If the wormhole moved in that plane,
Eq.~\eqref{eq:f1-ns5-actions} would apply: the instanton charged under
its axion would become actionless when the angle swept there reached
$\pi/2$. Our wormhole moves only in the planes of $a$ and $b$, but we
can still assign to every root the angle
$\varphi_\alpha=\alpha(u_s,\epsilon u_v)$, the combination of $u_s$ and
$u_v$ fixed by its weights. The elementary instantons have
$\varphi=u_s$ and $u_v$, which never reach $\pi/2$. We observe that the
walls found above are governed by these angles, and this answers the
four questions in order.

\emph{(i) One wall for two branes.} The outer roots have
$\varphi=(u_s\pm3\epsilon u_v)/2$, and on a regular wormhole the larger
of the two exceeds every other angle. Equation~\eqref{eq:root-distance}
is the statement that it reaches $\pi/2$: the wall is a quarter turn of
the wormhole along an outer root. This suggests that the wall belongs to
the scalar manifold rather than to either brane, which would explain why
both branes meet it at the same place. It is an observation about the
explicit actions of Sec.~\ref{sec:root-wall}, not an independent
derivation: we have not
constructed the complexified isometry that would carry the wormhole
into the hyperbolic plane of an outer root. The inequalities
$|\varphi_\alpha|<\pi/2$ are those that define the complex crown of a
symmetric space~\cite{Gindikin2001}, so the walls are consistent with
lying on its boundary.

\emph{(ii) The sign.} The reflection $R$ exchanges the two outer roots.
The root system is symmetric under it, so the wall cannot depend on
$\epsilon$. The set of BPS charges is not, since real flux induces
string charge of one orientation only, so the brane must. The reversed
branes of footnote~\ref{fn:reversed} have their walls where the other
outer angle, of size $(3u_v-u_s)/2$, reaches $\pi/2$.

\emph{(iii) The complex field.} The D5 of Sec.~\ref{sec:orig-d5} is the
instanton of the outer root $(\tfrac12,\tfrac32)$ itself, but its own
axion, the period of $C_6$, vanishes on the wormhole. The phase
$3u_v/2$ of its action is the part carried by $b$ of either outer-root
angle, $(u_s\pm3u_v)/2$, and it misses the part $u_s/2$. Its wall
is therefore later, and since no axion does work on it, its action
turns instead of vanishing.

\emph{(iv) Equal angles.} The picture also gives one number for how
exposed a wormhole is. Start from the simplest case, a single
hyperbolic plane of radius $R$ probed by its own instanton. The angle
advances as $u=\tau/R$, the instanton is actionless at $u=\pi/2$, and
the wall is at $\tau_*=\pi R/2$. The smaller the radius, the sooner the
wall. For any wormhole, and any collection of instantons, let $\tau_*$
be the first wall along the path and define
\begin{equation}
 h_{\rm eff}\equiv\frac{\pi}{2\tau_*} .
 \label{eq:heff-def}
\end{equation}
This is the inverse radius of the single hyperbolic plane whose wall would lie at
the same distance, so a large $h_{\rm eff}$ means an early wall. In four
dimensions the wormhole ends at $\tau_{\rm end}=\pi\sqrt3/2$, which is
where a hyperbolic plane of radius $\sqrt3$ has its wall, so a wall precedes the
end exactly when
\begin{equation}
  h_{\rm eff}>\frac{\pi}{2\tau_{\rm end}}
 =\sqrt{\frac{D-2}{2(D-1)}}=\frac1{\sqrt3}\quad(D=4).
 \label{eq:heff}
\end{equation}
Each root $\alpha$ sweeps its angle at the rate
$\alpha(p_s,\epsilon p_v/\sqrt3)$, which is the projection of the unit
tangent $(p_s,\epsilon p_v)$ of the path on that root in the
orthonormal coordinates of Fig.~\ref{fig:root-angles}. So
$h_{\rm eff}$ is the largest such projection among the roots whose
instantons are available, that is, the support function of their convex
hull in the direction of the path. With the elementary instantons alone
the hull is the rhombus of Fig.~\ref{fig:root-angles}, and the largest
projection is $\max(p_s,p_v/\sqrt3)<1/\sqrt3$ on every regular
wormhole: there is no wall before the end. With all the roots the hull
is the hexagon, and $h_{\rm eff}=(p_s+\sqrt3\,p_v)/2>\sqrt3/2$. The two
hulls are extremal in the same direction. The equal-angle division
minimizes the elementary support, at $1/2$, and it points exactly along
a long root, where the hexagon has a vertex and $h_{\rm eff}=1$. The
wormhole that hides best from the elementary instantons is the one most
exposed to the bound states.

\subsection{Consequences of the wall}\label{sec:consequences}

The action of a brane along the wormhole answers a different question for
each of the two boundary conditions of Sec.~\ref{sec:wormhole}.

\paragraph{Fixed boundary values.} With the scalars fixed at the ends, at
values continued away from the real slice, as we have assumed so far, a
wall means that an instanton the low-energy theory leaves out
is no longer suppressed. In ten dimensions the D-instanton
contributions to the $R^4$ coefficient diverge where the D-instanton
becomes actionless~\cite{Maldacena2026}. Here the corresponding
quantity is the scalar metric itself. The K\"ahler moduli, the dilaton
and their axions belong to hypermultiplets, and their kinetic terms are
the metric on which the wormhole is built. A wrapped fivebrane has four
universal fermion zero modes, which two scalar insertions absorb, so it
corrects this metric at the one-instanton
level~\cite{Alexandrov2021}. Corrections to the kinetic terms are what
Ref.~\cite{Licciardello2026} expects from the instantons that enforce
the bound when the degree exceeds three.

Take the universal axion that the wormhole does not move,
$\tilde\sigma$, dual to the spacetime two-form $C_{\mu\nu}$ (the NS
two-form $\tilde B_{\mu\nu}$ of the S-dual variables). Its kinetic
term is the simplest one to examine, because the action $I_k$ of the
fluxed D5 does not depend on $\tilde\sigma$. For $|k|=t_0$ at equal
angles the one-instanton
correction to its kinetic coefficient
is~\cite[Eqs.~(2.8), (3.21)--(3.22)]{Alexandrov2021}
\begin{equation}
 \frac{|\delta g^{(1)}_{\tilde\sigma\tilde\sigma}|}{g^{\rm tree}_{\tilde\sigma\tilde\sigma}}
 =\frac{\tilde V}{\sqrt{2\pi U_k}}\,e^{-\Ree I_k}
 =\frac{\sqrt{\tilde V\tilde g_s}}{2^{7/4}\pi}\,e^{-\Ree I_k}.
 \label{eq:finite-observable}
\end{equation}
The cost $U_k$ stays positive at the wall, so the prefactor stays
finite, and it exceeds one when $\tilde V\tilde g_s>2^{7/2}\pi^2\approx
10^2$. A single wrapping then corrects the metric by as much as its
tree-level value while $\Ree I_k$ is still positive, before the wall is
reached.

The sum over wrappings is the four-dimensional counterpart of the
divergence of the $R^4$ coefficient. For a primitive charge, $m$ wrappings contribute in
proportion to $(\sigma_\gamma e^{-I_k})^m/\sqrt m$, with a fixed sign
$\sigma_\gamma$~\cite[Eq.~(3.16)]{Alexandrov2021}. When the phase per
wrapping is one, which a constant real part of an RR axion can arrange,
the sum grows as $\sqrt{\pi/\Ree I_k}$ and diverges at the wall. We do
not claim that the exact metric is singular there. The fluxes $+k$ and
$-k$ lose their suppression together, their charges are mutually
nonlocal, and the exact answer depends on their
interaction~\cite{Twistors}. What the calculation shows is that the
instanton expansion of the scalar metric, on which the wormhole is
built, has failed by the time the wall is reached.

Finite volume corrects the cost of the brane through the curvature and
the Euler characteristic of $X$ and through worldsheet instantons, by
relative amounts of order $1/t^2$, $\zeta(3)/t^3$ and $e^{-2\pi t}$,
and it corrects the path through the one-loop term of the scalar
metric. The action crosses zero with a slope of order one in units of
its cost, so these finite-volume corrections move the wall
only slightly. The wall is thus where the leading-order theory predicts
its own failure. Since the instanton correction to the metric is already
of order one somewhat earlier, the low-energy description fails at or
before $\tau_*$.

\paragraph{Fixed charges.} We now return to the second boundary
condition of Sec.~\ref{sec:wormhole}, in which the charges of the
wormhole are fixed. The question is then different: does inserting a brane lower the action? For a brane that couples to the axions through fixed
charges the answer is the action we have computed. Such a brane is a
source of the axions' flux. Placed at a point of the wormhole it
changes the charge carried through the throat, and to leading order the
total action changes by the brane's cost minus its charges times the
axions at that point, measured from the throat~\cite{Lin2026}. Its
equation of motion places it where this change is smallest. Where the change is negative, inserting the brane lowers the action, and the wormhole is not the leading saddle with its charges. This is the fragmentation of wormholes proposed in
\cite{VanRiet2020} and made quantitative in \cite{Lin2026}. The question is
one about flux sectors, and its answer needs no semiclassical
description of the scalars; for such branes this meets the caveat of
Ref.~\cite{Witten} quoted in Sec.~\ref{sec:wormhole}.

The D5 with flux couples through fixed charges, its D5 charge and the
anti-D1 charge $\kappa k^2/2$. Its action is negative just beyond its
wall, which lies before the end, so inserting it lowers the action of every regular opposite-sign wormhole of fixed charge.

The D5--NS5 is different. The internal axion does not multiply a fixed
charge of this brane; it enters its worldvolume field, so the string
charge the brane carries depends on the axion itself. On the wormhole
that charge is neither quantized nor real, and the argument with fixed charges does not apply to it. The anti-D5 with flux does couple through fixed charges, but its wall lies
before the end only when $p_s < (\sqrt{2} - 1/\sqrt{3})/2$
(Table~\ref{tab:branes}). For the remaining same-sign wormholes of fixed
charge, including the one at equal angles, whether and how they fragment
is therefore open.

\section{Many axions}\label{sec:many}

Could a wormhole escape by sharing its excursion among hundreds of
axions, each moving only a short way~\cite[Sec.~5.2.1]{Maldacena2026}?
Recall how sharing worked for two. Each axion spans a hyperbolic plane
with its saxion, and the instanton charged under it keeps a fraction
$1-\sin u$ of its cost [Eq.~\eqref{eq:f1-ns5-actions}], where the angle
$u$ is the distance travelled in that plane divided by its radius. Now
suppose that $N$ independent hyperbolic planes of unit radius shared the excursion
equally, each probed only by its own instanton. Every angle would be
$\tau/\sqrt N$, and for $N\ge4$ none would reach $\pi/2$ before the
end. This is the dilution familiar from
$N$-flation~\cite{Nflation,Rudelius2015,Brown2015}. With many axions the instantons would hardly notice the wormhole.

A Calabi--Yau threefold is not such a product. Its saxions are the
dilaton $s$ and the areas of its curves, and on the slice where the RR axions vanish, their kinetic terms follow
from the function $K=-\log s-\log V$, as in
Ref.~\cite{Licciardello2026}. The volume is cubic in the areas, so $sV$ has
degree four in the saxions, one from the dilaton and three from the
volume, however many areas there are. What counts is this degree and
not the number of axions~\cite{Licciardello2026,Etheredge2026}. When all
saxions grow in proportion, the sector moves as a single hyperbolic plane whose
radius is the square root of its
degree~\cite[Eqs.~(6.45), (6.49)]{Licciardello2026}. The volume alone,
of degree three, gives radius $\sqrt3$, and its angle reaches $\pi/2$
exactly at the end. Degree four gives radius $2$ and the angle
$\tau/2$, which stops short of $\pi/2$: this is the equal-angle
wormhole of Sec.~\ref{sec:trajectory}, and its gap is the gap on every
threefold. Adding axions neither dilutes the elementary instantons nor
gives a bound state a gain of $\sqrt N$; the degree sets both. For degree above three, Ref.~\cite{Licciardello2026} infers that
instantons outside the mutually supersymmetric class must take over. On
the two-axion wormhole these are the fivebranes of
Sec.~\ref{sec:fivebranes}. Do they still cut the wormhole off when all
the internal axions move?

The K\"ahler class $J$
assigns to each curve $C$ its area $C\cdot J$, and the internal
displacement is the class $\beta=\Imm B_{\rm int}$, so that a string on
$C$ receives work $2\pi|C\cdot\beta|$. On the ray,
$\beta=(\Imm b)H$ and every curve sees the same ratio
$|C\cdot\beta|/(C\cdot J)=|d_b|$, the internal component of $\bd$ in
Eq.~\eqref{eq:zY}; off the ray each curve sees its own. Distances are measured with the
kinetic metric $G_J=-\partial_J^2\log V$ of the internal sector, which
on $J=vH$ gives the second hyperbolic plane of Eq.~\eqref{eq:metric}. A wormhole
is regular if all its saxions stay finite through the end.

\subsection{On the K\"ahler ray}\label{sec:ray}

Let the areas of all curves grow in proportion, $J=vH$, with the axion
along the same class, $\beta\propto H$, for any integral class $H$ that
gives every curve a positive area. On this K\"ahler ray
Sec.~\ref{sec:fivebranes} carries over unchanged, for two reasons. The
wormhole stays on the ray: the force on the saxions is the gradient of
$\log V$, which points along $J$ itself, so nothing pushes the shape of
$X$ away from $H$. And the brane does not see the rest of the
threefold: a flux proportional to the K\"ahler form is uniform over
$X$, so it minimizes the DBI action on any Calabi--Yau metric, and the
intersection number drops out of every action. Hence on every
threefold, for every regular division of the excursion, the D5 with
flux $|k|=t_0$ on the opposite-sign wormhole and the D5--NS5 bound
state with $p/q=g_{s,0}$ on the same-sign one have their walls at
$\tau_*=\pi/(p_s+\sqrt3\,p_v)$ of Eq.~\eqref{eq:root-distance}. The
homogeneous wormhole of Ref.~\cite{Licciardello2026} is the equal-angle
same-sign member of this family.

A product makes the contrast explicit. When the volume
factorizes as $v_1v_2v_3$, the dilaton and the three areas move on four
independent hyperbolic planes of unit radius. This is the dilution argument with $N=4$: each elementary instanton sees only its own hyperbolic plane, and if
the excursion is shared equally none of them reaches $\pi/2$ before the
end. The D5 with flux in each factor sees all four. Its fivebrane charge
feels the external axion, and the strings induced along each factor
feel that factor's axion. Take the signs of the four axions for which
the work adds over all these charges, the analogue of the
opposite-sign wormhole. With the flux in each factor equal to that
factor's area at the throat, as in Sec.~\ref{sec:dual-d5}, the action
of the brane vanishes where half the sum of the four angles reaches
$\pi/2$, for every regular division of the excursion, and no other flux
does better. Sharing among more hyperbolic planes does not help the wormhole,
because the bound state adds the shares up. The wall lies between
$\pi/2$ and $\pi/\sqrt3$, the same range as on the K\"ahler ray, and the
brane is the counterpart of the string that winds diagonally across
$T^2\times T^2$~\cite{Maldacena2026}.

\subsection{When the threefold changes shape}\label{sec:offray}

Away from the ray the wormhole changes the shape of $X$: curves grow at
different rates, and the internal axion no longer points along the
K\"ahler class. Two questions return. Do the elementary instantons now
fare better or worse? And do the fivebranes still catch the wormhole?

\paragraph{Strings always see the displacement.} A string measures the
whole displacement along its curve and not one component of it. Let
$\rho=\sup_C|C\cdot\beta|/(C\cdot J)$ be the best ratio of work to cost
among the strings. Then
\begin{equation}
 G_J(\beta,\beta)\le3\rho^2 :
 \label{eq:hull}
\end{equation}
the length of any displacement is at most $\sqrt3$ times what the
best-placed string sees of it, for every $h^{1,1}$. The reason is
short. The classes $J\pm\beta/\rho$ are nonnegative on every curve, and
expanding the cubic volume in them writes $3-G_J(\beta,\beta)/\rho^2$
as a variance (Appendix~\ref{app:hull}).

In a frame $e_i$ orthonormal for $G_J$, the charge per unit cost of a
string on $C$ is the vector $\bz_C=(C\cdot e_i)/(C\cdot J)$, and
Eq.~\eqref{eq:hull} says that in every direction some string, of one
orientation or the other, has a component of at least $1/\sqrt3$. The
hull of these vectors therefore contains the ball of radius
$1/\sqrt3$, the inverse radius of a sector of degree
three.\footnote{This is the normalization of Sec.~\ref{sec:geometry}, in
which the string of the ray has length $1/\sqrt3$. The radius is $\sqrt{2/3}$ in the normalization of
Ref.~\cite{Etheredge2026}, whose scalar metric is half of ours.} This
is the convex-hull form of the axion weak gravity conjecture for the
K\"ahler axions. Reference~\cite{Etheredge2026} derives it for the
wrapped strings from the weak gravity conjecture for particles in five
dimensions, and asks whether bounds of this kind extend to hull
conditions over the whole charge lattice. For the K\"ahler axions of
every Calabi--Yau threefold, Eq.~\eqref{eq:hull} shows that it follows
from the cubic volume alone. With the dilaton and the NS5 included, the
ball has radius $1/2$, the inverse radius of degree four, which is less
than the $1/\sqrt3$ that a wall before the end requires
[Eq.~\eqref{eq:heff}]. On the ray the elementary instantons have exactly
this radius, so changing the shape never shrinks their hull below its
size on the ray. But the ray already hides well enough: the elementary
instantons alone cannot guarantee a wall, for any $h^{1,1}$.

\paragraph{The bound states follow the shape.} The charges of a
fivebrane are quantized, so a D5 with flux $kH$ keeps the same integers
along the path while the K\"ahler form drifts away from the direction
of $H$. Both fivebranes are BPS bound states, and a BPS state does not
disappear under a small change of the moduli. For the D5 with flux, the
flux rearranges itself over the deformed threefold, and the cost remains
the modulus of the central charge.\footnote{For the D5 with flux the supersymmetric
configuration solves the deformed Hermitian Yang--Mills
equation~\cite{Marino1999,Jacob2014}. On the ray it is
$F=(k/t)\tilde J$, and near the ray solutions exist on any threefold,
since the subsolution condition of Ref.~\cite{Pingali2019} holds there.}
The D5--NS5 has no flux to rearrange: its worldvolume field is the
two-form of the background, and its action depends smoothly on the
shape. On the ray both actions cross zero at their walls with a nonzero slope.
For wormholes near the ray they therefore still cross zero, close to the
walls of the ray, on every threefold and for both relative signs. The
mechanism of Sec.~\ref{sec:fivebranes} is not an artifact of the frozen
shape. Each step of this argument, however, holds only near the ray.
Under a finite change of shape, a supersymmetric brane with the same
charges need not exist, for instance because the phases of the
constituents come into line at a wall of marginal stability, where the
bound state splits; and the zero of the action can move past the end.
What remains open is therefore wormholes far from the ray, which share
the internal excursion unequally among the K\"ahler axions and change
the shape of $X$ by a finite amount.

\section{Discussion}\label{sec:discussion}

A wormhole can hide from every instanton charged under one axion, but
not from their bound states. On the two-axion wormhole of a
Calabi--Yau threefold a fivebrane carrying string charge becomes
actionless at $\tau_*=\pi/(p_s+\sqrt3\,p_v)$, between $1/\sqrt3$ and
$2/3$ of the way to the end, for every regular division of the
excursion and for either relative sign of the axions. The sign changes
the brane but not the wall: it is a D5 with magnetic flux on the
opposite-sign wormhole and a bound state of D5 and NS5 branes on the
same-sign one. We expect this to be the generic situation with several
axions: the wormhole is a family, and the choices that label it decide
which brane serves and where, though not whether one does. The breakdown
does not wait for the wall. When $\tilde V\tilde g_s\gtrsim10^2$ a
single wrapping of the fluxed D5 already corrects a kinetic term by as
much as its tree-level value, and at the wall the sum over wrappings can
diverge, which is the four-dimensional counterpart of the divergence of
the $R^4$ coefficient. With many axions the count that matters is the
degree, four, and not $h^{1,1}$. On the K\"ahler ray of every threefold
the same branes cut the wormhole off at the same walls, and near the ray
they still do, for both signs; far from the ray the question is open.

Three things explain these results. Which brane enforces the bound is
decided by the phases of central charges: the work of the axions adds
over the charges of a brane while its cost does not, and where real flux
cannot help, the continued two-form supplies an imaginary worldvolume
field. Many axions do not help because the degree, not the number of
axions, sets the geometry, while the bound states persist when the shape
of the threefold changes slightly, because they are BPS. And the wall
lies, we observe, where the wormhole completes a quarter turn along a
root of the full scalar manifold, in a direction that the wormhole never
excites; the branes that reach it carry charge under axions that vanish
on the wormhole. A truncation to the axions that move sees neither. The
inequalities that locate the wall are those that define the complex
crown of a symmetric space, the domain to which the eigenfunctions of
its invariant differential operators extend
holomorphically~\cite{Kroetz2005,Kroetz2008}. One such eigenfunction is
the Eisenstein series that gives the $R^4$ coefficient in ten
dimensions, and for imaginary $C_0$ the crown of
$SL(2,\mathbb R)/SO(2)$ is the region $|\Imm C_0|<1/g_s$, exactly where
the D-instanton keeps its action~\cite{Maldacena2026}. This suggests
that the imaginary distance bound is the statement that a wormhole must
leave the crown of its full scalar manifold before it reaches its end.

With the charges of the wormhole fixed instead of its boundary values,
the same actions say whether inserting a brane lowers the action. For the opposite 
sign the D5 with flux does, on every regular division. For the same sign the question is open. At
equal angles the same-sign wormhole is the homogeneous wormhole of
Ref.~\cite{Licciardello2026}. Its favored charges are mutually
supersymmetric, and it is the case that resists both linearly coupled
branes and the argument with fixed charges.

The results also bear on the weak gravity conjecture. The imaginary
distance bound and the convex-hull form of the axion weak gravity
conjecture are met by the same objects. The instantons that enforce the
bound are not exotic non-BPS objects but BPS bound states that are not
mutually supersymmetric with the elementary ones, regular branes of the
weakly coupled theory. On the homogeneous wormhole of
Ref.~\cite{Licciardello2026} the instanton that becomes actionless is
the D5--NS5 bound state. On its reflection, the opposite-sign wormhole,
it is the fluxed D5, which obeys the bound of that reference on the
action per unit charge. Mutual supersymmetry is therefore the wrong
organizing principle. Reference~\cite{Etheredge2026} argues for a bound
on the total degree of a theory by assuming that instantons lighter than
the mutually supersymmetric ones do not arise. Ours arise and are
regular, so this assumption cannot be taken for granted, although the
bound itself is not contradicted at degree four. The same reference also notes, as
unexplained, that for degree above three a wormhole has more charge per
unit action than the supersymmetric instantons. In the opposite-sign
direction the fluxed D5 has more than the wormhole; in the same-sign
direction, where that puzzle is posed, the comparison at fixed charge is
open. Finally, the hull condition bounds the effective decay constant of
the K\"ahler axions in every direction, so alignment and
$N$-flation~\cite{Nflation,Rudelius2015,Brown2015} get no help from
these axions at large volume.

Two broader consequences are conjectures. If the same bound states exist
in the $\mathcal N=1$ compactifications of phenomenology, estimates of
how strongly wormholes break the shift symmetry of an axion, which bear
on the axion quality problem~\cite{Hebecker2018}, should start from the wormhole with the brane inserted rather than from the wormhole alone.
And that wormholes are cut off, or fragment, fits the expectation that a
consistent theory of quantum gravity has no free
$\alpha$-parameters~\cite{ColemanHerrings,McNamara2020}.

All of this is at leading order at large volume, in type IIB with
$\mathcal N=2$ supersymmetry and in either weakly coupled regime, with
the RR axions and the complex structure held fixed and a
reflection-symmetric throat. Several problems remain. Far from the
K\"ahler ray we have not shown that any brane cuts the wormhole off; we
conjecture that every regular wormhole is cut off at least as early as
on the ray, $h_{\rm eff}\ge\sqrt3/2$. Whether same-sign wormholes of
fixed charge fragment is open. With fixed boundary values, whether any
saddle remains is the question the imaginary distance bound was
designed to raise. For the Calabi--Yau wormholes studied here, the
low-energy theory does not get to answer it: the bound states of its
instantons cut the wormhole off first.

\acknowledgments
I am grateful for several useful conversations with Jan de Boer. Research at ICTS-TIFR is supported by the Department of Atomic Energy, Government of India, under Project Identification No. RTI4001. AI agents, including Astra and Claude Opus, have been used for literature review, cross-checks, and editing; the author has verified all results and takes full responsibility for the paper.

\appendix
\section{Bounds on the work of smooth magnetic flux}\label{app:flux-bounds}

We prove the bounds used in Secs.~\ref{sec:dual-d5}
and~\ref{sec:reversal}. At equal angles in the R--R regime, with
$u=\tau/2$, no single D-brane with a smooth Abelian magnetic field has
$W/U$ above $\sqrt2\sin u$ on the opposite-sign wormhole, or above
$\sin u$ on the same-sign wormhole, at the order of the DBI action
without derivative corrections. On the opposite-sign wormhole the
imaginary potentials can be represented by
\begin{equation}
 \Imm\tilde C_6=\frac{\sin u}{\tilde g_s}\,\frac{\tilde J^3}{6},\qquad
 \Imm\tilde C_2=-\frac{\sin u}{\tilde g_s}\,\tilde J ,
\end{equation}
which have the periods $\Imm a=s\sin u$ and
$\Imm b=-(t/\tilde g_s)\sin u$; on the same-sign wormhole $\tilde C_2$
has the opposite sign.

\emph{Opposite sign.} At a point on a D5, a rotation of the tangent
space brings any real $F$ to block form, with strengths
$\lambda_i=\tan\alpha_i\ge0$ in three orthogonal two-planes, and it
leaves the DBI density unchanged. The DBI density is
$\prod_i\sqrt{1+\lambda_i^2}$ times that of the bare brane. Flux in two
of the planes induces D1 charge along the third. In
$\tilde J\wedge F^2/2$ only the component of $\tilde J$ along that third
plane enters, and since $\tilde J$ evaluated on an orthonormal pair of
vectors is at most one in magnitude, each induced term in the work density is at most
$\lambda_i\lambda_j$ in the same units. The local ratio of work to cost
is therefore at most $\sin u$ times
\begin{equation}
 \frac{1+\sum_{i<j}\lambda_i\lambda_j}
      {\prod_i\sqrt{1+\lambda_i^2}}
 =\cos\alpha_3\cos(\alpha_1-\alpha_2)
 +\sin\alpha_3\sin(\alpha_1+\alpha_2)
 \le\cos\alpha_3+\sin\alpha_3\le\sqrt2 .
 \label{eq:bound}
\end{equation}
The inequality survives integration over the brane, it holds for any
real $F$, including fields with a $(2,0)$ component, and
$F=\pm\tilde J$ attains it everywhere. On a D3 the same argument gives
$(\lambda_1+\lambda_2)/\prod_{i=1}^2\sqrt{1+\lambda_i^2}
=\sin(\alpha_1+\alpha_2)\le1$. For a D1 the integral of $\tilde J$ is
bounded by the wrapped area, and a D($-1$) couples to neither potential.

\emph{Same sign.} In units of $2\pi\sin u/\tilde g_s$ the work density
on a D5 is $\tilde J^3/6-\tilde J\wedge F^2/2$, the real part of
$(\tilde J+iF)^3/6$. For a field of type $(1,1)$, with signed strengths
$\lambda_i=\tan\alpha_i$ in three orthogonal complex tangent planes,
this is $\Ree\prod_i(1+i\lambda_i)=1-\lambda_1\lambda_2
-\lambda_2\lambda_3-\lambda_3\lambda_1$ times the density on the bare
D5, and the ratio of work to cost is $\sin u$ times
\begin{equation}
 \frac{|1-\lambda_1\lambda_2-\lambda_2\lambda_3-\lambda_3\lambda_1|}
      {\prod_i\sqrt{1+\lambda_i^2}}
 =\bigl|\cos(\alpha_1+\alpha_2+\alpha_3)\bigr|\le1 .
 \label{eq:reverse-angles}
\end{equation}
For a general real $F$, in an oriented orthonormal frame with axes along
the magnetic planes of $F$, $(\tilde J+iF)^3/6$ is the Pfaffian
$\operatorname{Pf}(\tilde J+iF)$ times the volume form $\tilde J^3/6$.
Each column of $\tilde J+iF$ has norm $\sqrt{1+\lambda_i^2}$, whether or
not $\tilde J$ is block-diagonal in this frame: its real part
$\tilde Je_a$ is a unit vector and its imaginary part $Fe_a$ has length
$\lambda_i$. Hadamard's bound on a
determinant, with $\operatorname{Pf}(\tilde J+iF)^2=\det(\tilde J+iF)$,
gives
\begin{equation}
 |\Ree\operatorname{Pf}(\tilde J+iF)|
 \le|\operatorname{Pf}(\tilde J+iF)|
 \le\prod_i\sqrt{1+\lambda_i^2},
 \label{eq:reverse-general}
\end{equation}
whose right-hand side is the DBI density. Integrating,
$|W|/U\le\sin u<1$ for every smooth Abelian magnetic field, including
fields with a $(2,0)$ component.

\section{The hull inequality}\label{app:hull}

Products of classes denote intersection numbers on $X$, $J$ is a
K\"ahler class, $V=J^3/6$ and $G_J=-\partial_J^2\log V$. For a class
$\beta$ we write
\begin{equation}
 L\equiv G_J(\beta,\beta)
 =\Bigl(\frac{3J^2\beta}{J^3}\Bigr)^2-\frac{6J\beta^2}{J^3}.
\end{equation}
Suppose first that $J+\beta$ and $J-\beta$ are nonnegative on every
curve, and put $J_\pm=(J\pm\beta)/2$. The mixed intersection numbers of
such classes are nonnegative, so the four numbers
\begin{equation}
 m_i=\binom3i\,\frac{J_+^{\,i}J_-^{\,3-i}}{J^3},\qquad i=0,\dots,3,
\end{equation}
are nonnegative weights, and they sum to one because $J_++J_-=J$.
Since $\beta=J_+-J_-$,
\begin{equation}
 \frac{(J+z\beta)^3}{J^3}=\sum_{i=0}^3m_i\,(1+z)^i(1-z)^{3-i}.
\end{equation}
At $z=0$ the first two derivatives of the polynomial multiplying $m_i$
are $2i-3$ and $(2i-3)^2-3$. Taking minus the second derivative of the
logarithm gives
\begin{equation}
 L=3-4\operatorname{Var}_m(i)\le3 ,
 \label{eq:L-variance}
\end{equation}
where the variance is that of $i$ under the weights $m_i$.

For a general $\beta\ne0$, let $\rho$ be the supremum of
$|C\cdot\beta|/(C\cdot J)$ over curves, which is finite and positive
because $J$ is a K\"ahler class. The classes $J\pm\beta/\rho$ are
nonnegative on every curve, and Eq.~\eqref{eq:L-variance} applied to
$\beta/\rho$ is Eq.~\eqref{eq:hull}. A related bound on real axion
ranges uses the same metric~\cite{Rudelius}. In a frame orthonormal for
the full metric, a unit vector with component $\cos\theta$ along the
external axion and $\sin\theta$ along $\beta$ has projection
$\cos\theta$ on the NS5 and $\rho\sin\theta/\sqrt L\ge\sin\theta/\sqrt3$
on the best string. The larger of the two is at least $1/2$, which is
the radius quoted in Sec.~\ref{sec:offray} for the sector with the
dilaton.

\end{document}